\documentclass[superscriptaddress,preprintnumbers,amsmath,amssymb,prd,nofootinbib,preprint]{revtex4-1}
\pdfoutput=1
\usepackage{graphicx}
\usepackage{epstopdf}
\usepackage{dcolumn}
\usepackage{bm}
\usepackage{hyperref}
\usepackage{color}
\usepackage{amsmath}
\usepackage{cancel}
\usepackage{xpatch}

\newenvironment{minilinespace}{ \baselineskip = 0.65cm }

\begin{document}


\def\a{\alpha}
\def\b{\beta}
\def\c{\varepsilon}
\def\d{\delta}
\def\e{\epsilon}
\def\f{\phi}
\def\g{\gamma}
\def\h{\theta}
\def\k{\kappa}
\def\l{\lambda}
\def\m{\mu}
\def\n{\nu}
\def\p{\psi}
\def\q{\partial}
\def\r{\rho}
\def\s{\sigma}
\def\t{\tau}
\def\u{\upsilon}
\def\v{\varphi}
\def\w{\omega}
\def\x{\xi}
\def\y{\eta}
\def\z{\zeta}
\def\D{\Delta}
\def\G{\Gamma}
\def\H{\Theta}
\def\L{\Lambda}
\def\F{\Phi}
\def\P{\Psi}
\def\S{\Sigma}

\def\o{\over}
\def\beq{\begin{align}}
\def\eeq{\end{align}}
\newcommand{\gsim}{ \mathop{}_{\textstyle \sim}^{\textstyle >} }
\newcommand{\lsim}{ \mathop{}_{\textstyle \sim}^{\textstyle <} }
\newcommand{\vev}[1]{ \left\langle {#1} \right\rangle }
\newcommand{\bra}[1]{ \langle {#1} | }
\newcommand{\ket}[1]{ | {#1} \rangle }
\newcommand{\EV}{ {\rm eV} }
\newcommand{\KEV}{ {\rm keV} }
\newcommand{\MEV}{ {\rm MeV} }
\newcommand{\GEV}{ {\rm GeV} }
\newcommand{\TEV}{ {\rm TeV} }
\newcommand{\1}{\mbox{1}\hspace{-0.25em}\mbox{l}}
\newcommand{\headline}[1]{\noindent{\bf #1}}
\def\diag{\mathop{\rm diag}\nolimits}
\def\Spin{\mathop{\rm Spin}}
\def\SO{\mathop{\rm SO}}
\def\O{\mathop{\rm O}}
\def\SU{\mathop{\rm SU}}
\def\U{\mathop{\rm U}}
\def\Sp{\mathop{\rm Sp}}
\def\SL{\mathop{\rm SL}}
\def\tr{\mathop{\rm tr}}
\def\mpl{M_{\rm Pl}}

\def\IJMP{Int.~J.~Mod.~Phys. }
\def\MPL{Mod.~Phys.~Lett. }
\def\NP{Nucl.~Phys. }
\def\PL{Phys.~Lett. }
\def\PR{Phys.~Rev. }
\def\PRL{Phys.~Rev.~Lett. }
\def\PTP{Prog.~Theor.~Phys. }
\def\ZP{Z.~Phys. }

\def\dd{\mathrm{d}}
\def\ff{\mathrm{f}}
\def\BH{{\rm BH}}
\def\inf{{\rm inf}}
\def\ev{{\rm evap}}
\def\eq{{\rm eq}}
\def\SM{{\rm sm}}
\def\Mpl{M_{\rm Pl}}
\def\GeV{{\rm GeV}}
\newcommand{\Red}[1]{\textcolor{red}{#1}}
\newcommand{\TL}[1]{\textcolor{blue}{\bf TL: #1}}

\title{
Higgs Parity Grand Unification
}

\author{Lawrence J. Hall}
\affiliation{Department of Physics, University of California, Berkeley, California 94720, USA}
\affiliation{Theoretical Physics Group, Lawrence Berkeley National Laboratory, Berkeley, California 94720, USA}
\author{Keisuke Harigaya}
\affiliation{School of Natural Sciences, Institute for Advanced Study, Princeton, New Jersey 08540, USA}

\begin{abstract}
The vanishing of the Higgs quartic coupling of the Standard Model at high energies may be explained by spontaneous breaking of Higgs Parity. Taking Higgs Parity to originate from the Left-Right symmetry of the $SO(10)$ gauge group, leads to a new scheme for precision gauge coupling unification that is consistent with proton decay.  We compute the relevant running of couplings and threshold corrections to allow a precise correlation among Standard Model parameters.  The scheme has a built-in solution for obtaining a realistic value for $m_b/m_\tau$, which further improves the precision from gauge coupling unification, allowing the QCD coupling constant to be predicted to the level of 1\% or, alternatively, the top quark mass to 0.2\%. Future measurements of these parameters may significantly constrain the detailed structure of the theory. We also study an $SO(10)$ embedding of quark and lepton masses, showing how large neutrino mixing is compatible with small quark mixing, and predict a normal neutrino mass  hierarchy. The strong CP problem may be explained by combining Higgs Parity with space-time parity.
\end{abstract}

\date{\today}

\maketitle

\begin{minilinespace}
\tableofcontents
\end{minilinespace}

\section{Introduction}

The discoveries of a perturbative Higgs boson at the Large Hadron Collider~\cite{Aad:2012tfa,Chatrchyan:2012xdj} and no new states beyond the Standard Model (SM) \cite{Aaboud:2017vwy,Sirunyan:2018vjp} suggest that the SM may be the correct effective theory of particle physics up to a scale orders of magnitude larger than the weak scale,
a possibility largely ignored before the Large Hadron Collider.  In such a scenario, progress in particle physics will depend on both precision measurements of SM parameters, as well as searches for rare processes, for example those violating baryon number, lepton numbers and CP.

Precision measurements can probe particle physics to extremely high energies.  In 1974 Georgi, Quinn and Weinberg proposed that measurements of the three gauge couplings of the SM, $g_{1,2,3}$, could test whether the three gauge forces of nature are unified into a single grand unified gauge force with coupling strength, $g_u$,  at some very high unified mass scale~$M_u$~\cite{Georgi:1974yf}.   The two fundamental UV parameters lead to a correlation among the three measured gauge couplings: $(\alpha_u, M_u) \rightarrow \{g_{1,2,3} \}$.  After decades of measurements, this correlation is at best a first order approximation, requiring very large threshold corrections from the unified scale to force the low energy gauge couplings to meet and to make $M_u$ sufficiently large to be consistent with the experimental limit on the proton lifetime.  Similarly, the simplest $SU(5)$~\cite{Georgi:1974sy} prediction for fermion masses, the ratio $m_b/m_\tau$~\cite{Chanowitz:1977ye}, is also at best a first order result, requiring large corrections. Nevertheless, unification is a bold and exciting vision that explains the gauge quantum numbers of the quarks and leptons, including charge quantization, and can be probed via precision measurements of SM parameters at low energy. 

Precision measurements of the weak mixing angle at LEP \cite{Decamp:1990ky} supported supersymmetric unification. Triggering the weak scale from supersymmetry breaking, $v \sim m_{susy}$, gave a successful correlation of the low energy gauge couplings via  $(g_u, M_u, m_{susy}/v \simeq 1) \rightarrow \{g_{1,2,3} \}$ \cite{Dimopoulos:1981yj,Dimopoulos:1981zb,Sakai:1981gr,Ibanez:1981yh,Einhorn:1981sx,Marciano:1981un}. While theories with a sufficiently long proton lifetime were easily constructed, the absence of superpartners at the Large Hadron Collider now makes it difficult to identify $m_{susy}$ with the weak scale, weakening the theoretical basis for this correlation.

\begin{figure}[t]
\includegraphics[width=0.7 \linewidth]{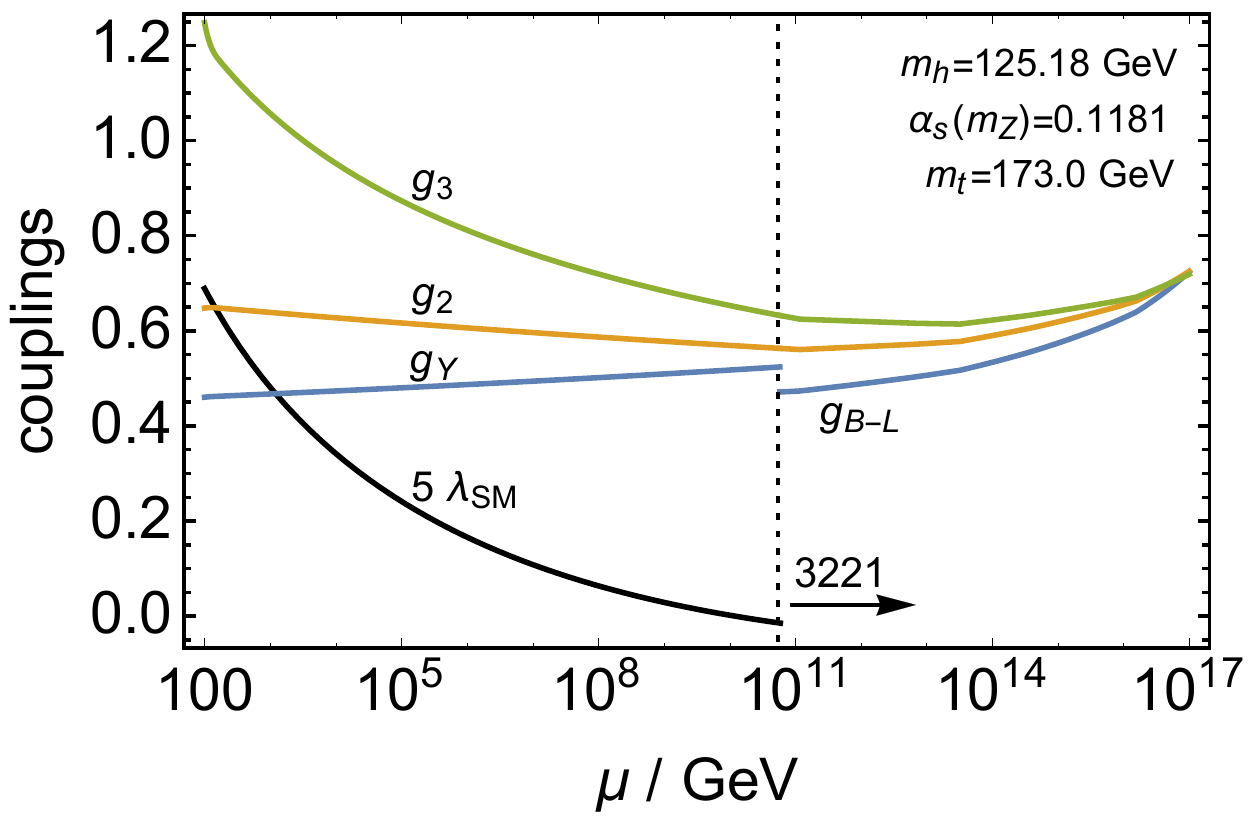}
\caption{Precise gauge coupling unification via Higgs Parity. The intermediate scale is the energy scale where the running Higgs quartic coupling of the Standard Model nearly vanishes.}
\label{fig:hpgut}
\end{figure}

With a 125 GeV Higgs and the SM valid to sufficiently high energies, the Higgs quartic coupling of the SM passes through zero at a scale of order $(10^9-10^{12})$ GeV~\cite{Buttazzo:2013uya}, as shown in Fig.~\ref{fig:hpgut}.  This very striking behavior suggests that new physics lies at the scale where the Higgs quartic coupling vanishes,  and that this new physics should explain the vanishing quartic via a new symmetry.  One possibility is that the new symmetry is supersymmetry; although the vanishing of the quartic is not guaranteed, it does occur in a large portion of parameter space \cite{Hall:2013eko,Hall:2014vga}.  We have recently introduced another possibility, ``Higgs Parity"~\cite{Hall:2018let}, that interchanges the weak $SU(2)$ gauge group (and SM Higgs, $H$) with a partner gauge group $SU(2)'$ (and partner Higgs, $H'$) 
\begin{align}
SU(2) \leftrightarrow SU(2)'  \hspace{1in} H(2,1) \leftrightarrow H'(1,2),
\end{align}
where the quantum numbers of $H$ and $H'$ refer to $(SU(2), SU(2)')$.  Spontaneously breaking $SU(2)'$ by $\vev{H'} = v'$ leads to the Higgs being a Nambu-Goldstone boson with $\lambda_{SM}(v') = 0$ at tree-level. Depending on the implementation, this can also solve the strong CP problem~\cite{Hall:2018let} and lead to interesting dark matter candidates~\cite{Dunsky:2019api}.  

\begin{figure}[t]
\includegraphics[width=0.45 \linewidth]{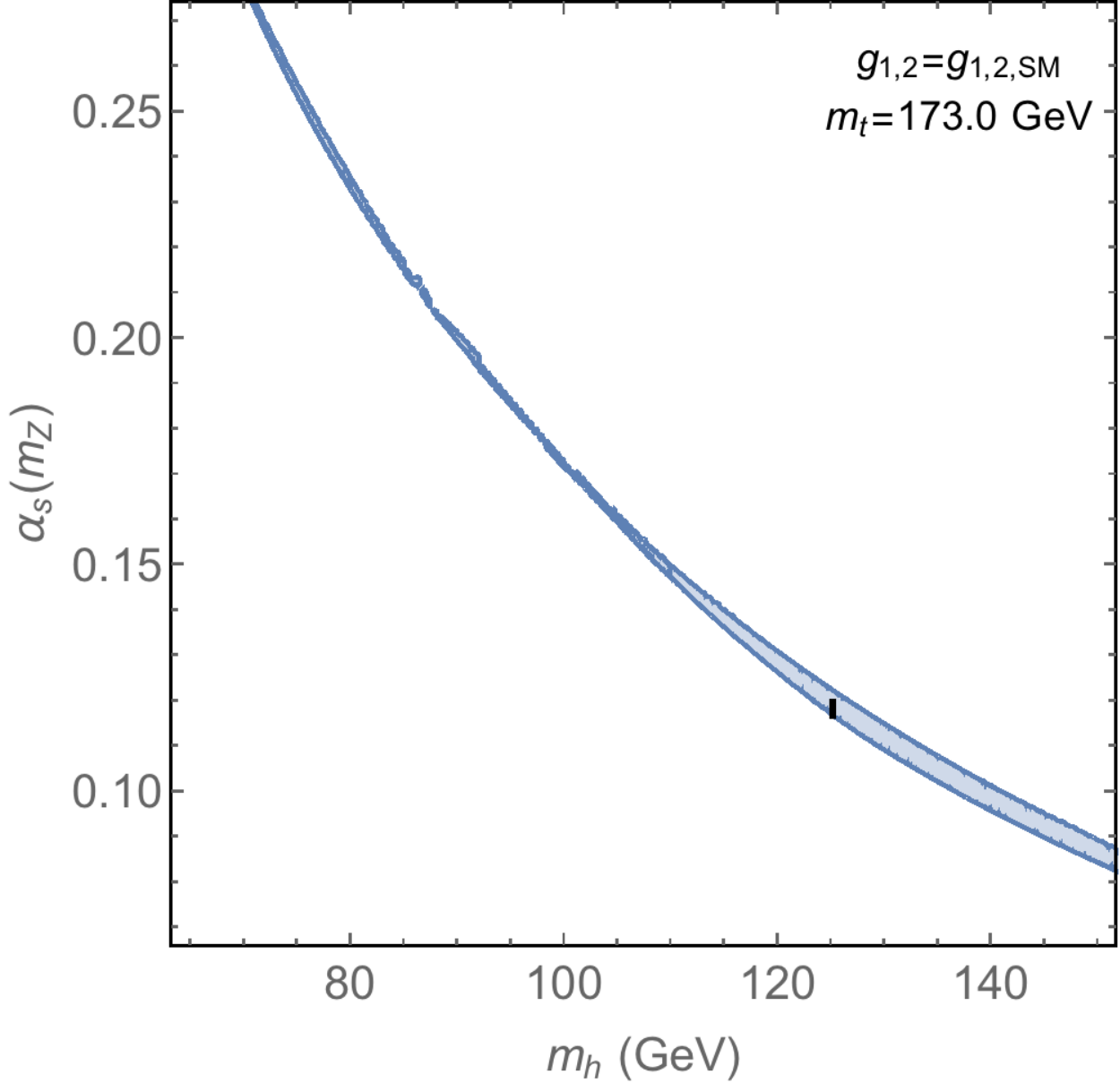}
\includegraphics[width=0.45 \linewidth]{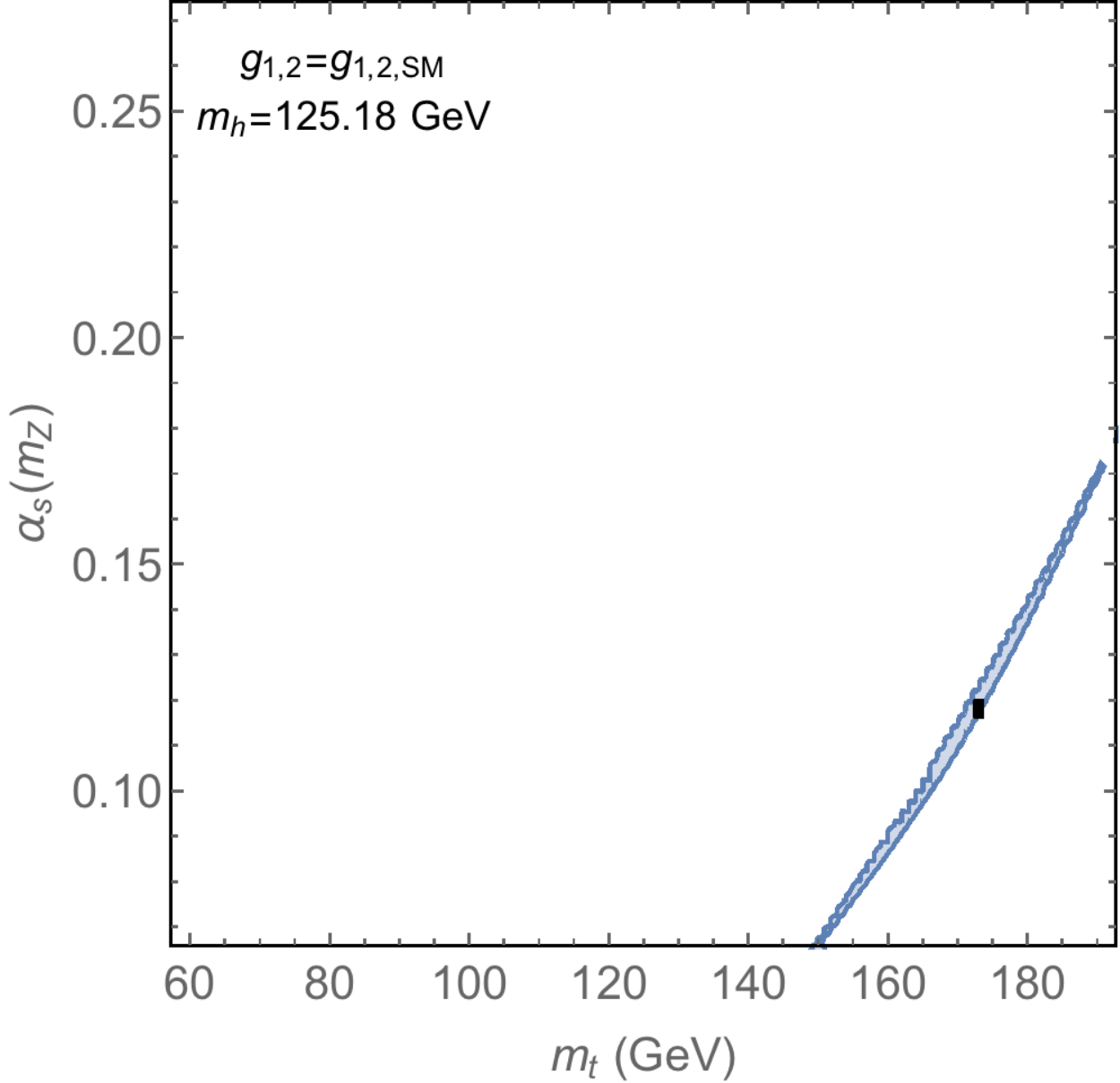}
\caption{Correlation of low energy parameters from coupling unification with Higgs Parity, projected into the $(m_h,\alpha_s)$ and $(m_t,\alpha_s)$ planes.}
\label{fig:global}
\end{figure}

In this paper we identify $SU(2) \times SU(2)'$ as the $SU(2)_L \times SU(2)_R$ subgroup of the unified $SO(10)$ gauge group~\cite{Georgi:1974my,Fritzsch:1974nn}, so that $v'$ is identified as the scale of parity breaking.  In $SO(10)$ unification, an intermediate scale of symmetry breaking introduces an extra free parameter so that the correlation of $\{g_{1,2,3} \}$ from gauge coupling unification is lost.  However, in theories with Higgs Parity, $v'$ is predicted from the Higgs mass so that a correlation is recovered, as illustrated in Fig.~\ref{fig:hpgut}; three parameters of the unified theory yield a correlation among four measured observables, $(g_u, M_u, v') \rightarrow \{g_{1,2,3} , m_h \}$.  In fact, the uncertainty in this correlation is dominated by the top quark Yukawa coupling $y_t$ via renormalization of the quartic coupling, so that in Higgs Parity Unification four UV parameters of the theory yield a correlation among five low energy observables~\cite{Hall:2018let}
\begin{align}
(g_u, M_u, y_t, v') \rightarrow \{g_{1,2,3} , m_h, m_t \}.
\end{align}
Fixing three of the observables to their central measured values, allows a projection of this correlation into a two-dimensional subspace, as shown for $(m_h, \alpha_s)$  and $(m_t, \alpha_s)$ in the left and right panels of Fig.~\ref{fig:global}. The blue shaded region allows for threshold corrections at the unification scale with $\Delta <10$ (see Eq.~(\ref{eq:Delta})). The black rectangles show the observed SM values. In Figs.~\ref{fig:hpgut} and \ref{fig:global} the gauge group above $v'$ is $SU(3) \times SU(2)_L \times SU(2)_R \times U(1)_{B-L}$.

The organization of the paper is as follows. Secs.~\ref{sec:HParity} and \ref{sec:HPGUT} summarize the essence of Higgs Parity unification.
In Sec.~\ref{sec:HParity}, we review how Higgs Parity explains the vanishing of the SM Higgs quartic at a high scale. Sec.~\ref{sec:HPGUT} discusses the embedding of Higgs Parity into $SO(10)$ unified theories and how gauge coupling unification is tied to the vanishing quartic coupling. 
Secs.~\ref{sec:GCU}-\ref{sec:pGUT} analyze the framework in more detail. Sec.~\ref{sec:GCU} examines the running of gauge couplings between electroweak and unified scales, including threshold corrections at the unification scale, and derives the Higgs Parity symmetry breaking scale required for successful precision gauge coupling unification. The generation of the SM fermion masses is discussed in Sec.~\ref{sec:yukawa}. We show how the $b/\tau$ mass ratio and the structure of neutrino masses arise from an $SO(10)$ unified theory. In Sec.~\ref{sec:scale}, we derive the threshold corrections to the SM Higgs quartic coupling at the Higgs Parity symmetry breaking scale, and show the relation between $(m_t, \alpha_s)$ and the Parity symmetry breaking scale. Finally, the prediction for $(m_t, \alpha_s)$ from the precise coupling unification is given in Sec.~\ref{sec:pGUT}.

\section{Higgs Quartic Coupling and Higgs Parity}
\label{sec:HParity}
In this section we review the relation between the nearly vanishing SM Higgs quartic coupling at high energy scales and the Higgs Parity symmetry breaking scale introduced in~\cite{Hall:2018let}. Consider a $Z_2$ symmetry which exchanges the SM $SU(2)$ gauge symmetry with a new gauge interaction $SU(2)'$, as well as the SM Higgs field $H(2,1)$ with its partner $H'(1,2)$. Here the brackets show the $SU(2)\times SU(2)'$ quantum numbers. We refer to this $Z_2$ symmetry as Higgs~Parity.

Well below the cut off scale, the following renormalizable scalar potential dominates the dynamics of $H$ and $H'$,
\begin{align}
V(H,H') = - m^2 (|H^2| + |H'|^2) + \lambda (|H|^2 + |H'|^2)^2 + \lambda' |H|^2 |H'|^2.
\label{eq:V}
\end{align}
We assume $m^2>0$ and $m \gg v$, the electroweak scale.  Higgs Parity is spontaneously broken by the vacuum expectation value (VEV)  $\vev{H'}\equiv v'$, with ${v'}^2 = m^2 / 2 \lambda$. After integrating out $H'$, the low energy effective potential of $H$ is 
\begin{align}
V_{LE}(H)= \lambda' v^2 |H|^2  - \lambda' ( 1 + \frac{\lambda'}{4\lambda})|H|^4.
\end{align}
To obtain the hierarchy $\vev{H} \ll v'$, it is necessary to take a very small value of $\lambda' \sim - v^2 / {v'}^2$, leading to a small value of the SM Higgs quartic coupling $\lambda_{\rm SM}\simeq 0$. This is the boundary condition on $\lambda_{SM}$ at the renormalization scale $\mu_c = v'$. Renormalization group running from the top quark yukawa makes $\lambda_{\rm SM}\simeq 0.1$ around the electroweak scale. From the IR perspective, the scale $v'$ is identified with the energy scale around which the SM Higgs quartic coupling vanishes. Threshold corrections to $\lambda_{\rm SM}(v')$ as well as a precise prediction for $v'$ are presented in Sec.~\ref{sec:scale}.

In this paper, we identify Higgs Parity with the Left-Right symmetry which can be embedded into $SO(10)$ grand unification. As we illustrated in the introduction and will elaborate in Sec.~\ref{sec:pGUT}, this identification leads to a non-trivial scheme for precise gauge coupling unification.

\section{Grand Unification and the Strong CP Problem}
\label{sec:HPGUT}

\subsection{Left-Right symmetry as Higgs Parity}
Let us first embed Higgs Parity into the Left-Right symmetry where $SU(2)'$ is identified with $SU(2)_R$. The gauge symmetry above the scale $v'$ is $SU(3)_c\times SU(2)_L\times SU(2)_R \times U(1)_{B-L}$ or $SU(4)\times SU(2)_L \times SU(2)_R$, which we refer to as $3221$ or $422$ for short.  422 is the Pati-Salam gauge group~\cite{Pati:1974yy}, and $SU(3)_c\times U(1)_{B-L}$ is a subgroup of $SU(4)$. The gauge quantum numbers of SM fermions, $H$ and $H'$ are shown in Table~\ref{tab:LR charge}.
The Left-Right symmetry, which we denote as $C_{LR}$, is
\begin{gather}
q \leftrightarrow \bar{q},~~ \ell \leftrightarrow \bar{\ell},~~H \leftrightarrow H', \nonumber \\
SU(2)_L \leftrightarrow SU(2)_R,~~~\text{charge conjugation on } SU(3)_c\times U(1)_{B-L} \text{ or } SU(4),
\end{gather}
and includes Higgs Parity. This results in the Higgs having gauge quantum numbers  identical to leptons, which is not standard for Left-Right theories~\cite{Beg:1978mt,Mohapatra:1978fy,Kibble:1982ae,Chang:1983fu,Chang:1984uy,Lazarides:1985my}.  The 3221 or 422 gauge groups are broken down to the $SU(3)_c\times SU(2)_L\times U(1)_Y$ group by the VEV of $H'$.

We may also combine Left-Right symmetry with another discrete $Z_2$ symmetry; the most interesting option being space-time parity,
\begin{gather}
q(t,x) \leftrightarrow i \sigma_2 \bar{q}^*(t,-x),~~\ell(t,x) \leftrightarrow i \sigma_2 \bar{\ell}^*(t,-x),~~H(t,x) \leftrightarrow {H'}^*(t,-x),\nonumber \\
SU(2)_L \leftrightarrow SU(2)_R,~~\text{parity transformation on gauge fields},
\end{gather}
which we denote as $P_{LR}$. As we will see, the strong CP problem may then be solved.

\begin{table}[htp]
\caption{The gauge charges of SM fermions, $H$ and $H'$ under 3221 or 422.}
\begin{center}
\begin{tabular}{|c|c|c|c|c|c|c|}
\hline
 & $q$ & $\ell$  & $(\bar{u}, \bar{d}) = \bar{q}$ &  $(N,\bar{e}) = \bar{\ell}$ & $H$ & $H'$ \\ \hline
$SU(3)_c$ & $3$ & 1  & $\bar{3}$ & 1 & 1& 1 \\
$SU(2)_L$ & $2$ & $2$ & $1$  & $1$ & $2$  & $1$ \\
$SU(2)_R$ & $1$ & $1$ & $2$ &  $2$ & $1$  & $2$ \\ 
$U(1)_{B-L}$ & $1/6$ & $-1/2$ & $-1/6$ &  $1/2$ & $-1/2$  & $1/2$ \\ \hline
$422$ &  \multicolumn{2}{c|}{$(4,2,1)$} &  \multicolumn{2}{c|}{$(\bar{4},1,2)$} & $(4,2,1)$ & $(\bar{4},1,2)$ \\ \hline
$SO(10)$ &  \multicolumn{4}{c|}{$16$} & \multicolumn{2}{c|}{$16$} \\ \hline  
\end{tabular}
\end{center}
\label{tab:LR charge}
\end{table}%

\subsection{Yukawa couplings and the strong CP problem}

The gauge charges in Table~\ref{tab:LR charge} forbid renormalizable yukawa couplings. Instead, the SM fermion masses arise from the mixing of $(f;\bar{f})=(q,\ell; \bar{q}, \bar{\ell})$ with extra massive fermions $(X;\bar{X})$ via yukawa couplings and masses of the form
\begin{gather}
[f_i \, x_{ij} \, \bar{X}_j \, H^{(\dag)} + \bar{f}_i \, x'_{ij} \, X_j \, H^{'(\dag)}] \hspace{0.25in} ~{\rm or}~ \hspace{0.25in} [f_i \, x_{ij} \,\bar{X}_j \, H^{'(\dag)} + \bar{f}_i \, x'_{ij} \, X_j \, H^{(\dag)}] \nonumber \\
+ \, m_{ij} X_i \bar{X}_j.
\label{eq:yukawa}
\end{gather}
After $H'$ obtains a VEV, $(f,\bar{f})$ mixes with $(X,\bar{X})$. A linear combination of them remains massless and has the yukawa coupling $ f_{SM_i} \, y_{ij} \,  \bar{f}_{SM_j} H^{(\dag)}$. If the mass of $X$ is much larger than $x v'$, we may integrate out $X$ to obtain a dimension-five operator $ f \bar{f} H^{(\dag)}H^{'(\dag)}$, which yields a yukawa coupling $y\sim x^2 v' / m$.

The strong CP problem can be solved by combining Left-Right symmetry with space-time parity, as the symmetry forbids the $\theta$ term and constrains the determinant of the quark mass matrix~\cite{Beg:1978mt,Mohapatra:1978fy}. See Refs.~\cite{Barr:1991qx,Kuchimanchi:1995rp,Mohapatra:1995xd,Mohapatra:1996vg,Mohapatra:1997su,Kuchimanchi:2010xs,DAgnolo:2015uqq,Albaid:2015axa,Babu:2018vrl,Mimura:2019yfi} for studies on Left-Right symmetric solutions to the strong CP problem. Refs.~\cite{Babu:1988mw,Babu:1989rb} propose a model with a structure for yukawa couplings  similar to ours and show that the strong CP problem is actually solved since $x'_{ij} = x_{ij}^*$ and $m_{ij}$ is Hermitian. They obtain the hierarchy $v \ll v'$ by softly breaking the Left-Right symmetry with space-time parity. In out setup the symmetry, which we call Higgs Parity, is spontaneously broken without soft breaking, predicting a vanishing $\lambda_{\rm SM}(v')$. 
Spontaneous breaking of parity generates a phase in the determinant of the quark mass matrix via two-loop quantum corrections~\cite{Hall:2018let}. Assuming that the couplings $x$ are $O(1)$, the corrections are safely below the current limit from the neutron electric dipole moment, but in the range that can be probed by planned experiments. The model of flavor presented in Sec.~\ref{sec:yukawa} does not obey this assumption, and the corrections may be larger.

\subsection{$SO(10)$ unification}
The 3221 and 422 theories can both be  embedded into $SO(10)$ grand unified theories. The $SO(10)$ gauge charges of the SM fermions, $H$ and $H'$ are shown in Table~\ref{tab:LR charge}. The SM fermions are unified into three ${\bf 16}$s, and the Higgs fields $H$ and $H'$ are also embedded into a ${\bf 16}$.

The symmetry breaking pattern is
\begin{align}
SO(10) \rightarrow
\begin{cases}
SU(3)\times SU(2)_L\times SU(2)_R\times U(1)_{B-L} \\
SU(4)\times SU(2)_L\times SU(2)_R
\end{cases}
\stackrel{H'}{\longrightarrow} \;\;\; SU(3)\times SU(2)_L \times U(1)_Y. \nonumber
\end{align}
The theory has three UV parameters relevant for gauge coupling unification: the $SO(10)$ gauge coupling, the $SO(10)$ symmetry breaking scale, and the LR symmetry breaking scale $v'$.
As there are also three SM gauge coupling constants, it is not surprising that one can typically find a set of the three UV parameters that allow coupling unification.
However, as we have shown, the LR symmetry breaking scale is not a free parameter when it is linked to Higgs Parity breaking, but is determined by the running of the SM Higgs quartic coupling. In this case, it would be significant if coupling unification were successful. In Fig.~\ref{fig:hpgut}, we fix the scale $v'$ using the central values of the Higgs mass, top quark mass and QCD coupling shown in the figure, and solve the RGE equations assuming the 3221 theory. Remarkably, gauge coupling unification occurs, and at a scale consistent with the proton lifetime.

In Sec.~\ref{sec:pGUT}, we analyze the precision of this coupling unification, including threshold corrections to gauge coupling constants at the unification scale, as well as the threshold corrections to the SM quartic coupling at the scale $v'$. The unification of the yukawa couplings is discussed in Sec~\ref{sec:yukawa}.

\subsection{Degree of fine-tuning}
We comment on the fine-tuning of parameters in the Higgs potential (\ref{eq:V}) required for symmetry breaking. First, $m^2$ must be fine-tuned by an amount $\Delta_{m^2}$, so that the Higgs Parity breaking scale $v'$ is much less than the cutoff scale $\Lambda$, which must be larger than the unified scale $M_u$. Secondly, $\lambda'$ must be fine-tuned by an amount $\Delta_{\lambda'}$ to obtain the electroweak scale $v$ from the scale $v'$. The total fine-tuning with Higgs Parity is the product
\begin{align}
\Delta_{HP} = \Delta_{m^2} \; \Delta_{\lambda'} = \frac{{v'}^2}{\Lambda^2} \times \frac{v^2}{{v'}^2} = \frac{v^2}{\Lambda^2},
\end{align}
which is independent of $v'$. This is because a smaller $v'$ requires more fine-tuning in $\Delta_{m^2}$, but this is compensated by less fine-tuning in $\Delta_{\lambda'}$ to obtain the electroweak scale from the scale $v'$.  It is important to note that Higgs Parity, $H\leftrightarrow H'$, ensures that the mass terms for $H$ and $H'$ in (\ref{eq:V}) are identical, so that the single fine-tune by $\Delta_{m^2}$ protects both scalars to the scale $v'$.  Given that the SM Higgs must be protected for electroweak symmetry breaking, there is no additional cost to protect $H'$: 
the smallness of the scale $v' \ll \Lambda$ requires no unnaturalness beyond that already needed for the weak scale. The total fine-tuning of the theory $\Delta_{HP} = v^2/\Lambda^2$ is nothing but the electroweak fine-tuning, which may be explained by environmental selection~\cite{Agrawal:1997gf,Hall:2014dfa}.

This is in contrast to the usual $SO(10)$ unification with an intermediate scale $v_I$. A smaller intermediate scale does not reduce the fine-tuning to obtain the electroweak scale, and hence the total fine-tuning is
\begin{align}
\Delta_{SO(10)} = \frac{v_I^2}{\Lambda^2} \times \frac{v^2}{\Lambda^2}.
\end{align}
This extra fine-tuning $v_I^2/ \Lambda^2$ cannot be explained by environmental selection of the electroweak scale, and requires an additional explanation.

\section{Gauge Coupling Unification and Parity Breaking Scale}
\label{sec:GCU}

We assume an $SO(10)$ gauge symmetry at a high energy scale, broken to $3221$ or $422$ at the unification scale. These are then broken to the SM gauge group by the VEV of $H'$.  One possibility is that Higgs Parity is $C_{LR}$, a $Z_2$ subgroup of $SO(10)$ that interchanges $SU(2)_L$ and $SU(2)_R$. In this case, the symmetry breaking chain and the required Higgs fields are
\begin{align}
SO(10) \hspace{0.2in} &\stackrel{\phi_{210}}{\longrightarrow} \hspace{0.2in} 3221 \times C_{LR} 
  \hspace{0.2in} \stackrel{H'}{\longrightarrow} \hspace{0.2in} SU(3) \times SU(2)_L \times U(1)_Y, \\
  SO(10) \hspace{0.2in} &\stackrel{\phi_{54}}{\longrightarrow} \hspace{0.2in} 422 \times C_{LR} 
  \hspace{0.2in} \stackrel{H'}{\longrightarrow} \hspace{0.2in} SU(3) \times SU(2)_L \times U(1)_Y.
\end{align}
To solve the strong CP problem, the symmetry to begin with is $SO(10) \times CP$. This symmetry is broken by the VEV of a field that is odd under both $C_{LR}$ and CP, so that the residual $Z_2$ symmetry for Higgs Parity is $C_{LR} * CP = P_{LR}$ and includes spacetime parity.  In this case 
\begin{align}
SO(10)\times CP \hspace{0.2in} &\stackrel{\phi_{45}^-}{\longrightarrow} \hspace{0.2in} 3221 \times P_{LR} 
  \hspace{0.2in} \stackrel{H'}{\longrightarrow} \hspace{0.2in} SU(3) \times SU(2)_L \times U(1)_Y, \\
  SO(10) \hspace{0.2in} &\stackrel{\phi_{210}^-}{\longrightarrow} \hspace{0.2in} 422 \times P_{LR}  \hspace{0.2in}  \stackrel{H'}{\longrightarrow} \hspace{0.2in} SU(3) \times SU(2)_L \times U(1)_Y.
\end{align}

In this section we compute the running of the gauge coupling constants from IR to UV, treating the parity symmetry breaking scale $v'$ as a free parameter.

Values of the SM gauge couplings derived from experiment are
\begin{align}
g_1(m_t) =0.4626, ~g_2(m_t) = 0.64779,~ g_3(m_t) =  1.1666
\end{align}
in the $\overline{\rm MS}$ scheme at a renormalization scale of $m_t$.
Here the hypercharge coupling is given in the normalization appropriate for grand unification and is called $g_1$, or occasionally $g_Y$ to avoid confusion with the $B-L$ gauge coupling.
Between the electroweak scale and the scale $v'$, the RGE equation at the two-loop level is given by
\begin{align}
\frac{\rm d}{{\rm dln} \mu}
\begin{pmatrix}
\frac{2\pi}{\alpha_1} \\
\frac{2\pi}{\alpha_2} \\
\frac{2\pi}{\alpha_3}
\end{pmatrix} = 
\begin{pmatrix}
- \frac{41}{10} \\
\frac{19}{6}  \\
 7
\end{pmatrix} + 
\begin{pmatrix}
- \frac{199}{100} & - \frac{27}{20}  & - \frac{22}{5} \\
 - \frac{9}{20}  & - \frac{35}{12} & -6 \\
- \frac{11}{20} & -\frac{9}{4} & 13
\end{pmatrix}
\begin{pmatrix}
\frac{\alpha_1}{2\pi} \\
\frac{\alpha_2}{2\pi} \\
\frac{\alpha_3}{2\pi}
\end{pmatrix}.
\end{align}

\subsection{$SU(3)\times SU(2)\times SU(2) \times U(1)$}
We match the $SU(3)\times SU(2)\times U(1)$ gauge coupling constants to those of 3221 at the $W'$ mass,
\begin{align}
\frac{2\pi}{\alpha_{Y} (m_{W'})} = \left( \frac{2}{5} \right) \frac{2 \pi}{\alpha_{B-L} (m_{W'})}  + \left( \frac{3}{5} \right) \frac{2\pi}{\alpha_{2} (m_{W'})}  - \frac{1}{10}, \hspace{0.4in} m_{W'} = \frac{g_2}{\sqrt{2}} v'.
\end{align}
Since $W'$ is the only heavy charged gauge boson at this scale, no mass-dependent threshold corrections are introduced from the gauge bosons.
The RGE equation in the 3221 theory is
\begin{align}
\frac{\rm d}{{\rm dln} \mu}
\begin{pmatrix}
\frac{2\pi}{\alpha_{1}} \\
\frac{2\pi}{\alpha_2} \\
\frac{2\pi}{\alpha_3}
\end{pmatrix} = 
\begin{pmatrix}
- \frac{9}{2} \\
\frac{19}{6}  \\
 7
\end{pmatrix} + 
\begin{pmatrix}
- \frac{23}{8} & - \frac{27}{4}  & - 2 \\
 - \frac{9}{8}  & - \frac{35}{12} & -6 \\
- \frac{1}{4} & -\frac{9}{2} & 13
\end{pmatrix}
\begin{pmatrix}
\frac{\alpha_{1}}{2\pi} \\
\frac{\alpha_2}{2\pi} \\
\frac{\alpha_3}{2\pi}
\end{pmatrix}.
\end{align}
Here we only show the contributions from gauge bosons, SM fermions, $H$ and $H'$; contributions from $X$ states are shown in Appendix \ref{sec:Xcontrib}.

We match the 3221 gauge couplings to that of $SO(10)$ at the mass, $M_{XY}$, of the XY gauge boson of charge $(3,2,2,-1/3)$. The only heavy gauge boson, other than the $XY$ gauge boson, has 3221 quantum numbers $(3,1,1,2/3)$. Taking this gauge boson to have mass $r_{XY} M_{XY}$, gives threshold corrections
\begin{align}
\frac{2\pi}{ \alpha_{1}(M_{XY})} &= \frac{2\pi}{ \alpha_{10}(M_{XY})} + 14\, {\rm ln} \, r_{XY} - \frac{4}{3} + \Delta_1  &\equiv    \frac{2\pi}{ \alpha_{10}(M_{XY})} + \Delta_{1,G} + \Delta_1, \nonumber \\
\frac{2\pi}{ \alpha_2(M_{XY})} &= \frac{2\pi}{ \alpha_{10}(M_{XY})} - 1 + \Delta_2 & \equiv    \frac{2\pi}{ \alpha_{10}(M_{XY})} + \Delta_{2,G}+ \Delta_2, \nonumber \\
\frac{2\pi}{ \alpha_3(M_{XY})} &= \frac{2\pi}{ \alpha_{10}(M_{XY})} + \frac{7}{2}\, {\rm ln} \, r_{XY} - \frac{5}{6} + \Delta_3& \equiv  \frac{2\pi}{ \alpha_{10}(M_{XY})} + \Delta_{3,G} + \Delta_3,
\end{align}
where $\Delta_i$ denote possible threshold corrections from scalars and fermions.
If the $SO(10)$ symmetry is broken by a VEV of ${\bf 45}$, $r_{XY}=2$. If it is broken by the VEV of the $SU(4)$ adjoint part of ${\bf 210}$, $r_{XY} = \sqrt{2}$. The VEV of ${\bf 54}$ or the $SU(4)$ singlet part of ${\bf 210}$ gives a mass only to the $XY$ gauge bosons, and makes $r_{XY}$ smaller.

\begin{figure}[t]
\includegraphics[width=0.45 \linewidth]{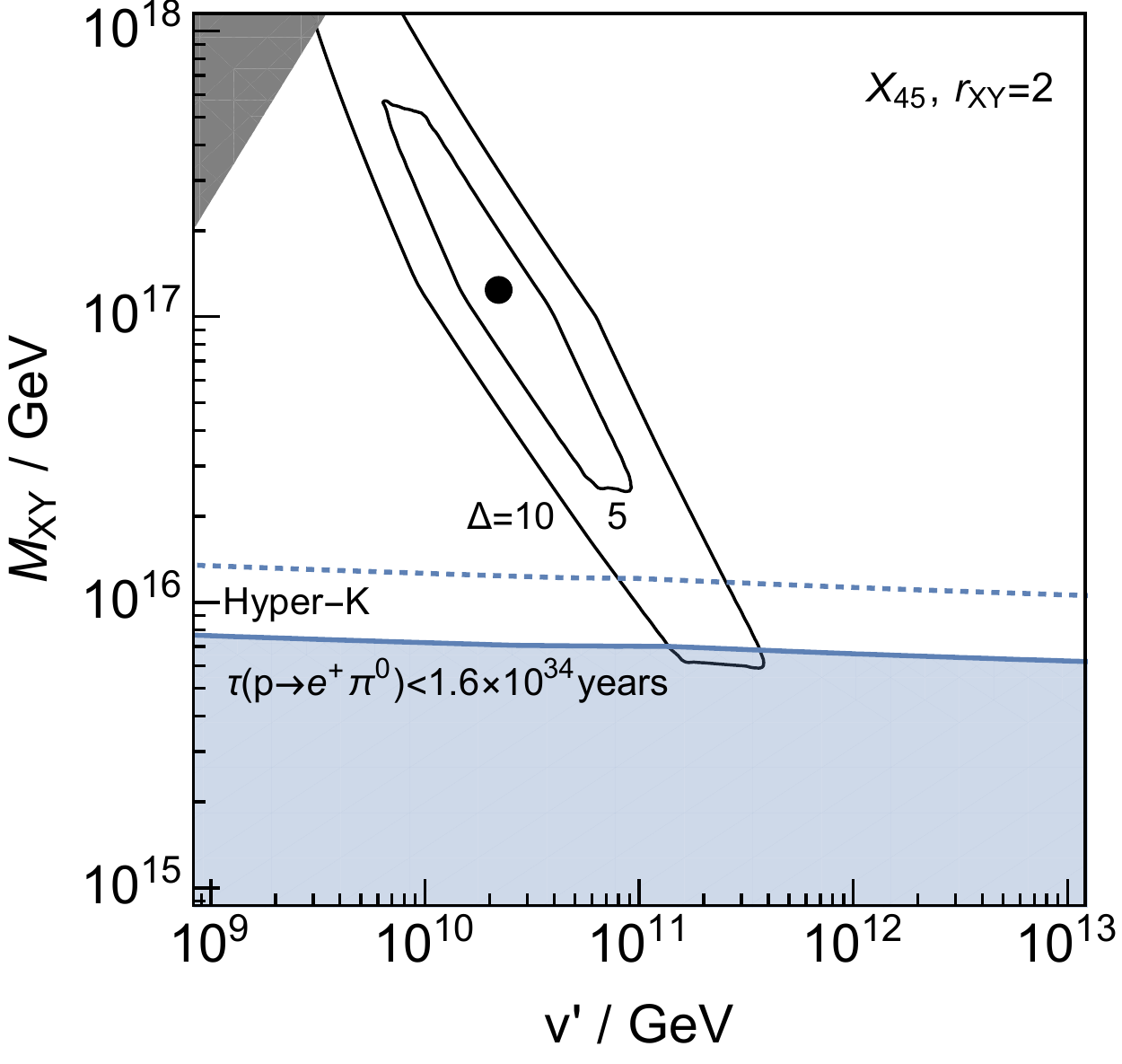}
\includegraphics[width=0.45 \linewidth]{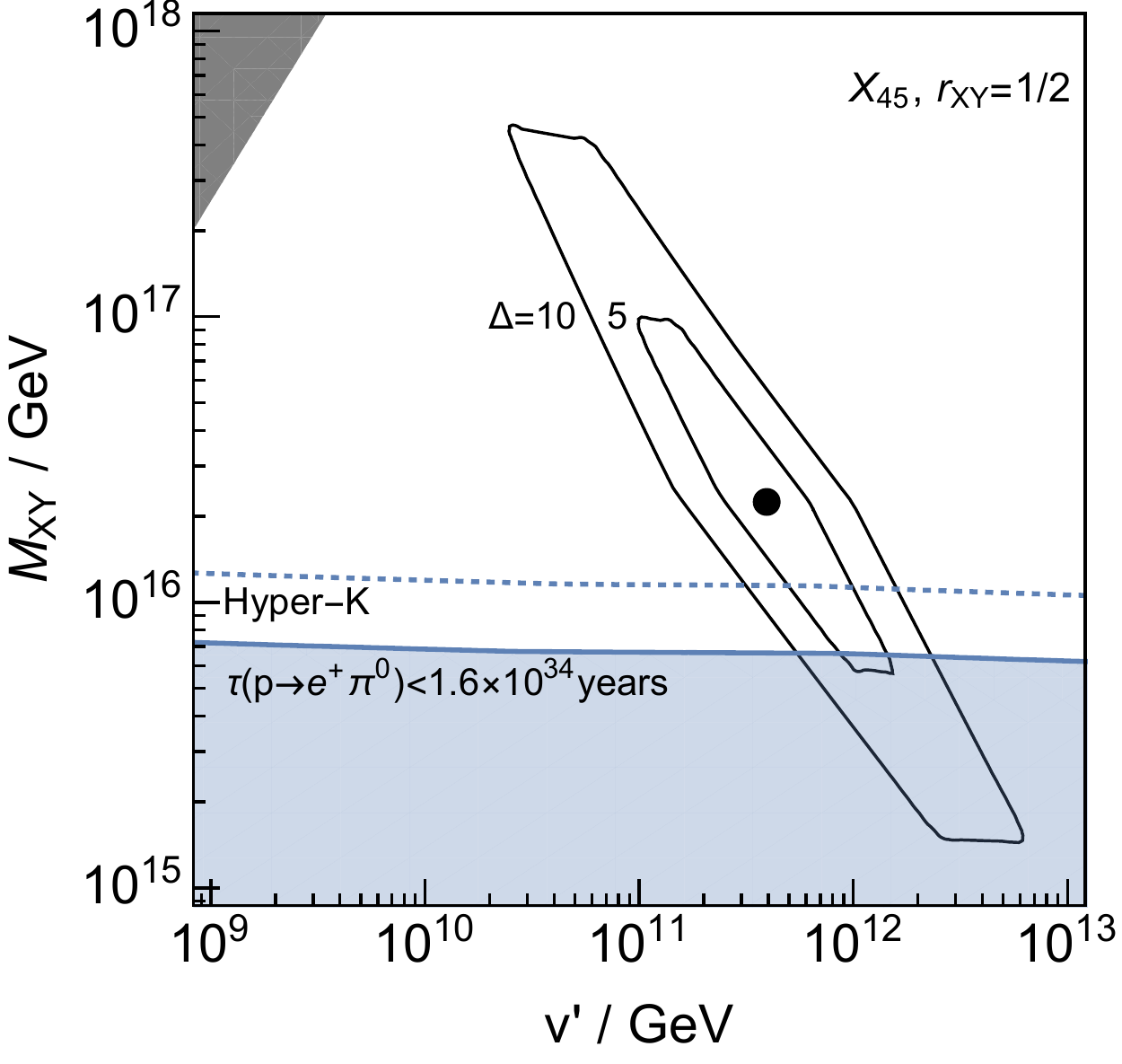}
\includegraphics[width=0.45 \linewidth]{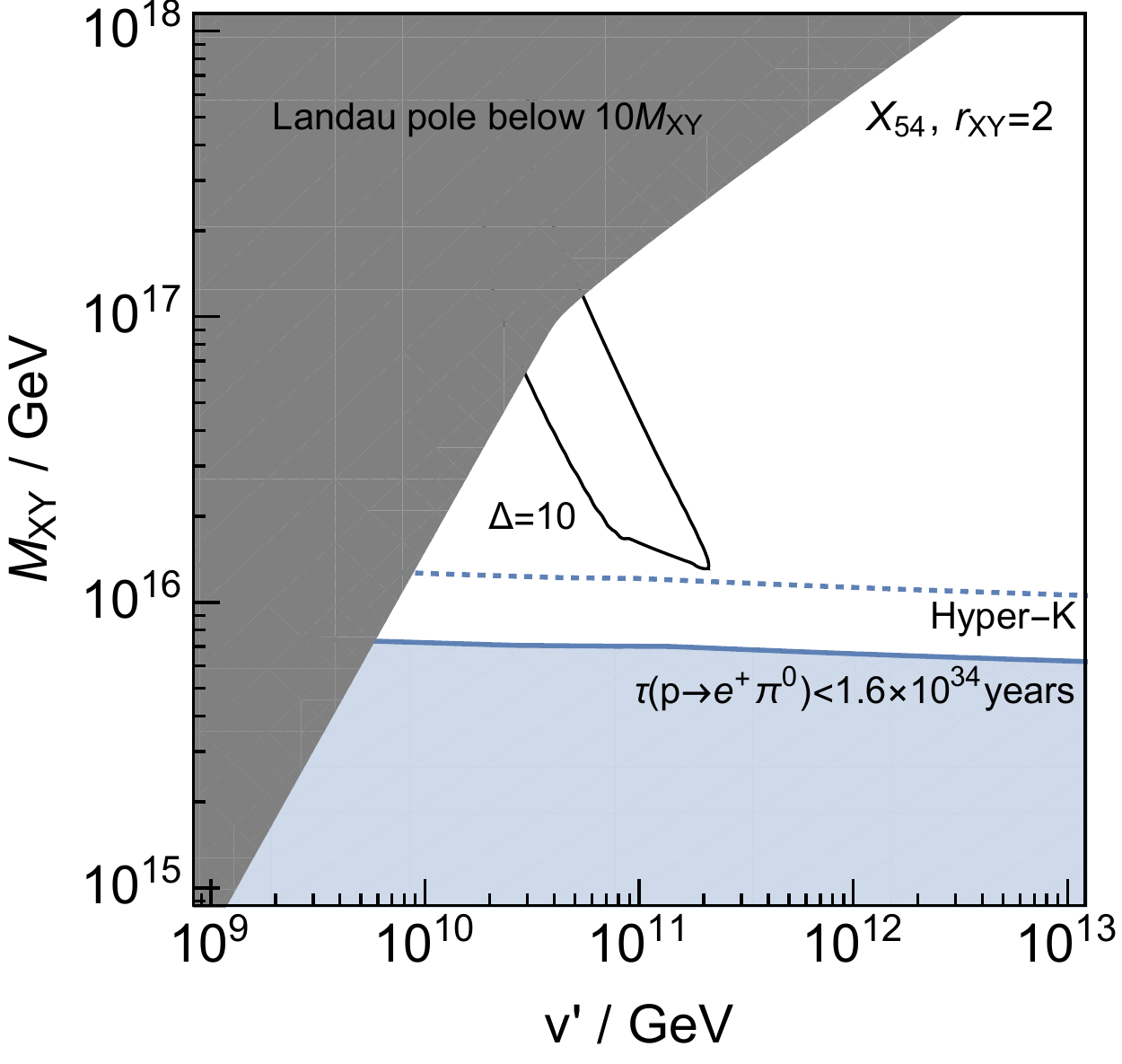}
\includegraphics[width=0.45 \linewidth]{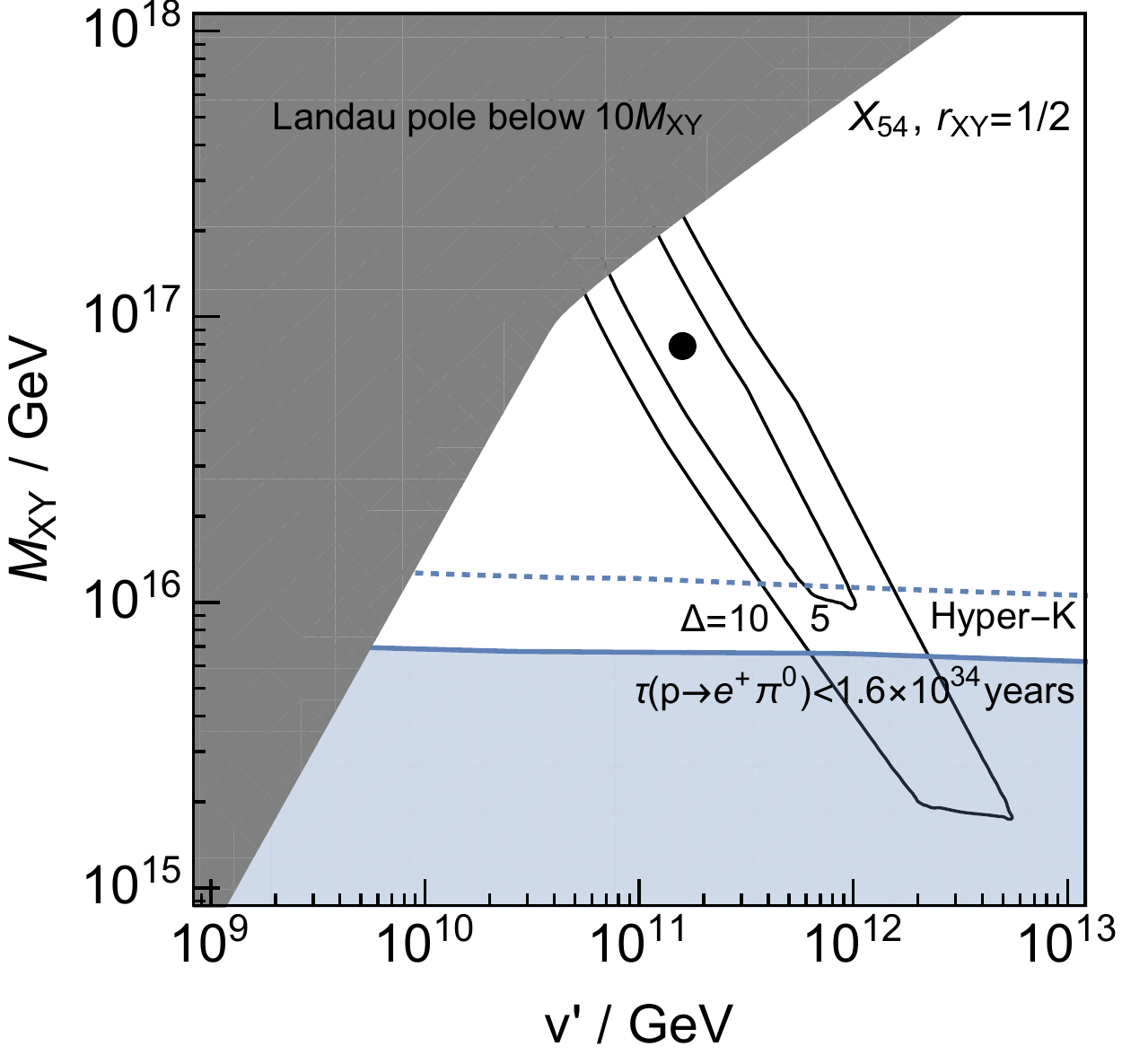}
\caption{Determination of $v'$ and $M_{XY}$ in the 3221 theory from gauge coupling unification alone.  The left (right) panels are for differing values of the unified gauge threshold corrections, and the contours show the effects of unified threshold corrections from scalars and fermions. The upper (lower) panels have the top yukawa coupling generated from the exchange of a ${\bf 45}$ (${\bf 54}$) $X$ state.}
\label{fig:3221_vp}
\end{figure}

For each $M_{XY}$, the threshold correction from scalars and fermions required for unification is
\begin{align}
\label{eq:Delta}
\Delta \; \equiv \; {\rm max}_{i,j}\left|\left( \frac{2\pi}{ \alpha_i} - \Delta_{i,G} \right) - \left( \frac{2\pi}{ \alpha_j} - \Delta_{j,G} \right) \right| 
\;=\; {\rm max}_{i,j}\left| \Delta_i -\Delta_j \right|.
\end{align}
In Fig.~\ref{fig:3221_vp}, we show contours of $\Delta$ in the $(v', M_X)$ plane, assuming $r_{XY}=2$ (left) and $1/2$ (right). The dot indicates the point where $\Delta=0$.  In the upper/lower panel, we assume that the $X$ multiplet generating the up yukawa couplings is ${\bf 45/54}$. We fix the $X$ masses so that the quark yukawa couplings are reproduced for $x = 1$. In the gray-shaded region, the Landau pole of the $SO(10)$ gauge coupling is less than $10 M_{XY}$, so that the precision of gauge coupling unification is spoiled. The blue-shaded region predicts too rapid a proton decay rate and is excluded by Super Kamiokande~\cite{Miura:2016krn}. The blue dotted line shows the sensitivity of Hyper Kamiokande~\cite{Abe:2018uyc}.

As $x$ is varied so the required value of $M_X$ changes.  However, in the case that the entire $SO(10)$ multiplet is degenerate, an order of magnitude change in $M_X$ only changes $\Delta$ by $\sim 1$, as this is a two loop effect.  The different 3221 irreducible representations, $X_a$, within a single $SO(10)$ multiplet receive non-degeneracies of only few 10\% or less from gauge radiative corrections below $M_{XY}$.  However, for successful flavor physics we allow order unity tree-level splittings between these masses leading to contributions to $\Delta$ of $(4/3) C \ln (M_a/M_b)$, where $C$ is a quadratic Casimir, normalized to 1/2 for a fundamental representation. Order unity splittings can give $\Delta \sim 1-3$, depending on the size and number of the $X$ multiplets.

In Appendix \ref{sec:SO10scalars} we compute contributions to $\Delta$ from scalar multiplets that break $SO(10)$. $\Delta$ is typically smaller than 1 if ${\bf 45}$ is the only such multiplet, while ${\bf 54}$ and ${\bf 210}$ multiplets allow for $\Delta$ of a few and 10, respectively.

Higher dimensional interactions between the $SO(10)$ symmetry breaking field and the gauge field in general split the gauge coupling constants at the unification scale. Assuming a suppression scale of the reduced Planck mass, splittings from a dimension five operator typically give $\Delta \simeq 10$ for a unification scale of $10^{17}$ GeV. In theories with CP symmetry at the unification scale, which solve the strong CP problem, the dimension five operator is forbidden, and the splittings from a dimension six operator typically give $\Delta \simeq 1$. At lower values of the unification scale these values of $\Delta$ are reduced.

\subsection{$SU(4)\times SU(2)\times SU(2)$}
We match the SM gauge coupling constants to those of the 422 theory at the $W'$ mass, 
\begin{align}
\frac{2\pi}{\alpha_3 (m_{W'})} =  \frac{2\pi}{\alpha_4 (m_{W'})}  + \frac{28}{5}{\rm ln} \frac{g_4}{g_2}- \frac{1}{6} , \hspace{0.7in} m_{W'} = \frac{g_2}{\sqrt{2}} v', \nonumber \\
\frac{2\pi}{\alpha_1 (m_{W'})} = \left(\frac{2}{5} \right) \frac{2\pi}{\alpha_4 (m_{W'})} + \left( \frac{3}{5} \right) \frac{2\pi}{\alpha_2 (m_{W'})} + \frac{21}{5}{\rm ln} \frac{g_4}{g_2} - \frac{7}{30}.
\end{align}
Since the values of $\alpha_4$ and $\alpha_2$ are known, the successful embedding of $U(1)_{Y}$ into the Pati-Salam gauge group fixes the scale $v'$. To take into account a possible threshold correction, we define
\begin{align}
\Delta_{422} \equiv \frac{2\pi}{\alpha_1 (m_{W'})} -  \frac{2}{5} \frac{2\pi}{\alpha_3 (m_{W'})} -   \frac{3}{5} \frac{2\pi}{\alpha_2 (m_{W'})} - \frac{21}{5}{\rm ln} \frac{g_4}{g_2} - \frac{3}{10}.
\end{align}

The RGE equation of the $SU(4)\times SU(2)\times SU(2)$ gauge coupling constants is 
\begin{align}
\frac{\rm d}{{\rm dln} \mu}
\begin{pmatrix}
\frac{2\pi}{\alpha_2} \\
\frac{2\pi}{\alpha_4}
\end{pmatrix} = 
\begin{pmatrix}
\frac{8}{3} \\
10  
\end{pmatrix} + 
\begin{pmatrix}
 - \frac{37}{6}  & - \frac{75}{4} \\
- \frac{15}{2} &  \frac{117}{4}  \\
\end{pmatrix}
\begin{pmatrix}
\frac{\alpha_2}{2\pi} \\
\frac{\alpha_4}{2\pi}
\end{pmatrix}.
\end{align}
Here we only show the contribution from the gauge bosons, the SM fermions, $H$ and $H'$. The contribution from the $X$ states is shown in Appendix \ref{sec:Xcontrib}.

\begin{figure}[t]
\includegraphics[width=0.5 \linewidth]{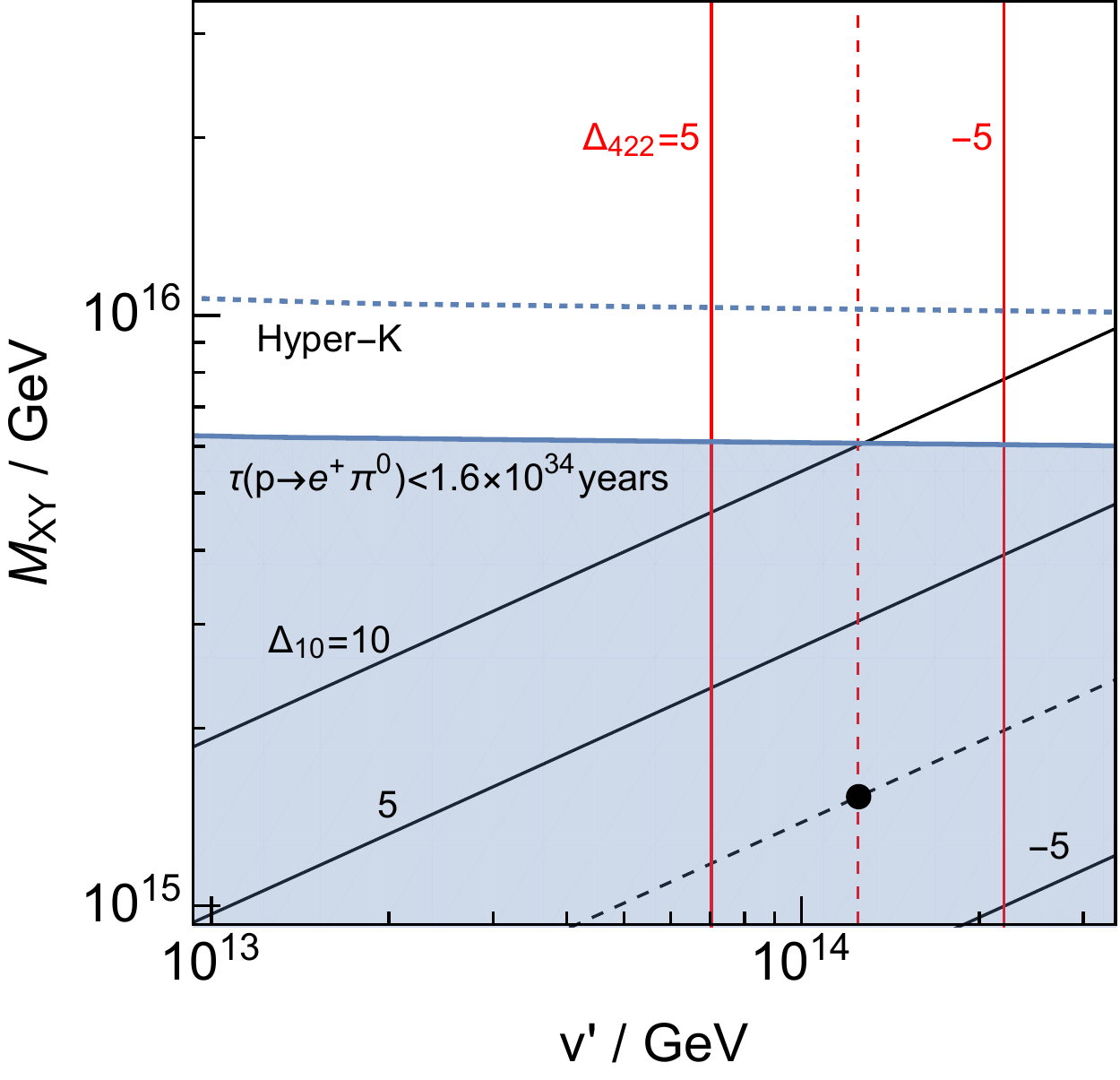}
\caption{Determination of $v'$ and $M_{XY}$ in the 422 theory from gauge coupling unification alone.  Contours of threshold corrections from scalars and fermions at the scale $v'$ ($\Delta_{422}$) and at the scale $M_{XY}$ ($\Delta_{10}$) are shown in red and black.}
\label{fig:422_vp}
\end{figure}

We match the 422 gauge couplings at the mass, $M_{XY}$, of the XY gauge boson, which is the only heavy gauge boson. The threshold corrections at $M_{XY}$ are
\begin{align}
\frac{2\pi}{ \alpha_2(M_{XY})} &= \frac{2\pi}{ \alpha_{10}(M_{XY})} - 1 + \Delta_2& \equiv    \frac{2\pi}{ \alpha_{10}(M_{XY})} + \Delta_{2,G} + \Delta_2, \nonumber \\
\frac{2\pi}{ \alpha_4(M_{XY})} &= \frac{2\pi}{ \alpha_{10}(M_{XY})} - \frac{2}{3}  + \Delta_4 & \equiv  \frac{2\pi}{ \alpha_{10}(M_{XY})} + \Delta_{4,G} + \Delta_4,
\end{align}
where $\Delta_{4,2}$ denote possible threshold corrections from scalars and fermions.
For each $M_{XY}$, we quantify the required value of the threshold correction by
\begin{align}
\Delta_{10} \; \equiv \; \left( \frac{2\pi}{ \alpha_4} - \Delta_{4,G} \right) - \left( \frac{2\pi}{ \alpha_2} - \Delta_{2,G} \right) \; = \; \Delta_4- \Delta_2~.
\end{align}
In Fig.~\ref{fig:422_vp}, we show the contours of $\Delta_{422}$ and $\Delta_{10}$. The parameter point where no threshold correction is required is already excluded by Super Kamiokande. A threshold correction of $\Delta_{10}\sim 10$ is necessary to evade the bound from proton decay. We estimate the typical magnitude of the threshold corrections from the unified scalar multiplets that break $SO(10)$ belonging to ${\bf 54}$ or ${\bf 210}$ in Appendix \ref{sec:SO10scalars}, and show that $\Delta_{10}$ is typically $O(1)$. This is because of the smallness of the contribution of scalar particles to the renormalization of gauge couplings. Threshold correction can be large if the theory near the unification scale is non-minimal; if the unified scale arises from the supersymmetry breaking scale, the threshold correction can be easily as large as 10.

\section{Yukawa Couplings}
\label{sec:yukawa}
The predictions from Higgs Parity coupling unification are affected by threshold corrections to $\lambda_{\rm SM}(v')$.  The SM yukawa couplings are generated from the mixing of $(q,\bar{q}, \ell,\bar{\ell})$ with the $X$ states when parity is broken by $H'=v'$. The leading correction to $\lambda_{\rm SM}(v')$ is expected to arise from the generation of the top quark yukawa coupling.  In this section we discuss how the SM yukawa couplings arise from the $SO(10)$ unified theory via interactions of (\ref{eq:yukawa}). We also show that there is a simple understanding of why the $b/\tau$ mass ratio deviates from the simplest expectation from grand unification, as well as why the neutrino masses and the mixings are not as hierarchical as those of quarks. We also comment on a possible impact on leptogenesis~\cite{Fukugita:1986hr}.

The $X$ states arise from ${\bf 45}, {\bf 54}$ or ${\bf 10}$ representations of $SO(10)$, whose decomposition into 3221 is shown in Table~\ref{tab:Xup}. ${\bf 45}$ and ${\bf 54}$ give up-type yukawa couplings and neutrino masses, while ${\bf 10}$ gives down-type quark and charged lepton yukawa couplings. We do not consider larger representations as they lead to the gauge couplings blowing-up below the unification scale. For complex 3221 representations, $Q,U$ and $D$, we omit their complex conjugates, $\bar{Q},\bar{U}$ and $\bar{D}$ from the table.  Non-singlet $SU(2)_R$ multiplets are decomposed into SM multiplets by giving the $U(1)_Y$ charge as a subscript; thus $Q$, which is an $SU(2)_R$ doublet, contains SM multiplets $(Q_{1/6}, Q_{-5/6})$. 

Terms in the Lagrangian of the $SO(10)$ theory that lead to quark and lepton masses are 
\begin{align}
{\cal L}_{SO(10)} = (\psi \, x_{45,54} \, X_{45,54}) \phi^\dag \, {\cal O}_G +  \frac{m_{45,54}}{2}  X_{45,54}^2  {\cal O}_G+
(\psi \,x_{10} \,X_{10}) \phi  \, {\cal O}_G +  \frac{m_{10}}{2}  X_{10}^2  {\cal O}_G 
\end{align}
where $\psi(q,\ell, \bar{q}, \bar{\ell})$ and  $\phi \supset H, H'$ are both in ${\bf 16}$, and ${\cal O}_G$ denotes possible insertions of fields with $SO(10)$ symmetry breaking vevs. Note $\psi X_{54} \phi^\dag$ is not an $SO(10)$ invariant, and hence requires a non-trivial ${\cal O}_G$. In the following we analyze the yukawa couplings in the 3221 theory. The discussion for the 422 theory is almost the same, as the 422 symmetry does not impose relations between the parameters in the 3221 theory except for one case that we mention below.  We study the generation of yukawa couplings in the up, down, charged lepton and neutrino sectors by integrating out the $X$ states.  
\begin{table}[htp]
\caption{Decomposition of $X$ states into representations of $3221$.}
\begin{center}
\begin{tabular}{|c|c|c|c|c|c|}
\hline
$SO(10)$ & \multicolumn{5}{c|}{{\bf 54}} \\ \hline
 & $Q$ & & & & $S$  \\ \hline
$SU(3)$ & 3 & 6 & 8  & 1 & 1 \\
$SU(2)$ & 2 & 1  & 1 & 3 & 1 \\
$SU(2)$ & 2 & 1 & 1 & 3  & 1 \\ 
$U(1)$ & $-1/3$ & $-2/3$ & 0 & 0 & 0  \\ \hline
\end{tabular}
\begin{tabular}{|c|c|c|c|c|c|c|}
\hline
$SO(10)$ & \multicolumn{6}{c|}{{\bf 45}} \\ \hline
 & $Q$ & $U$ & & $T_L$ & $T_R$ & $S$  \\ \hline
$SU(3)$ & 3 & 3 & 8  & 1 & 1 & 1 \\
$SU(2)$ & 2 & 1  & 1 & 3 & 1 & 1 \\
$SU(2)$ & 2 & 1 & 1 & 1 & 3 & 1 \\ 
$U(1)$ & $-1/3$ & 2/3 & 0 & 0 & 0 & 0 \\ \hline
\end{tabular}
\begin{tabular}{|c|c|c|}
\hline
$SO(10)$ & \multicolumn{2}{c|}{{\bf 10}} \\ \hline
 & $D$ & $\Delta$  \\ \hline
$SU(3)$ & 3 & 1  \\
$SU(2)$ & 1 & 2   \\
$SU(2)$ & 1 & 2  \\ 
$U(1)$ & $-1/3$ & $0$  \\ \hline
\end{tabular}
\end{center}
\label{tab:Xup}
\end{table}%

\subsection{Up-type quark yukawa couplings}

The $X$ states for the up-type yukawas couplings are in ${\bf 45}$ or ${\bf 54}$.
For ${\bf 54}$, the up yukawa couplings arise from the interaction and mass term
\begin{align}
\label{eq:uyukawa54}
{\cal L }_{54_Q} =   \bar{q} \, x_Q Q \, H^\dag   +  q \, x_Q \bar{Q} \, {H'}^\dag   + m_Q Q \bar{Q}.
\end{align}
Note that below the $SO(10)$ breaking scale, the $\psi (Q,U,...) \phi$ couplings $x_X$ and masses $m_X$ are given for each 3221 component of $X: (Q,U, D, ...)$. We allow these couplings and masses to deviate from strict $SO(10)$ relations by order unity amounts via ${\cal O}_G$. 

\begin{figure}[t]
\includegraphics[width=0.8\linewidth]{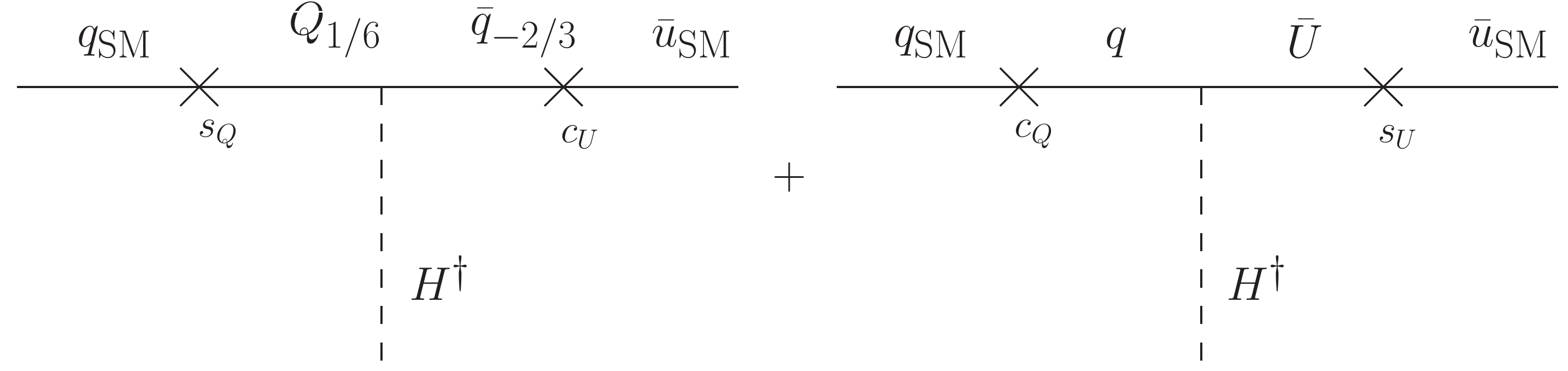}\\
\vspace{0.3cm}
\includegraphics[width=0.4\linewidth]{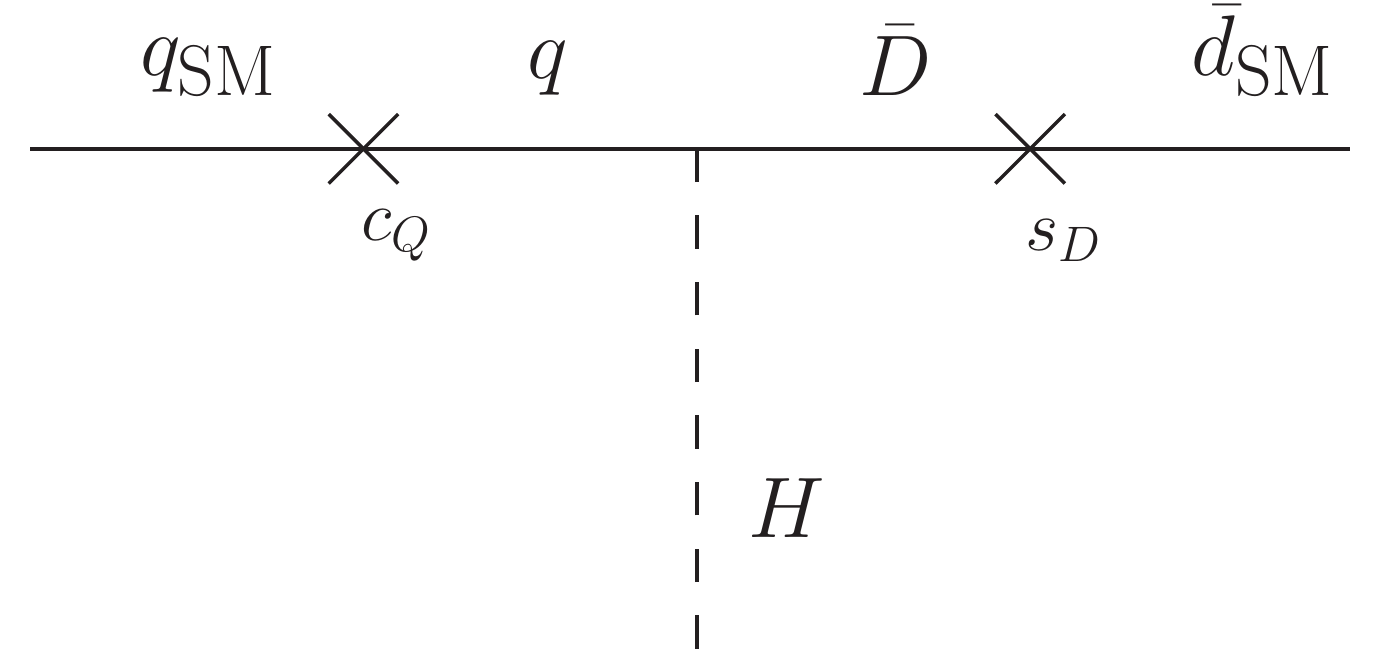}
\hspace{0.2cm}
\includegraphics[width=0.4\linewidth]{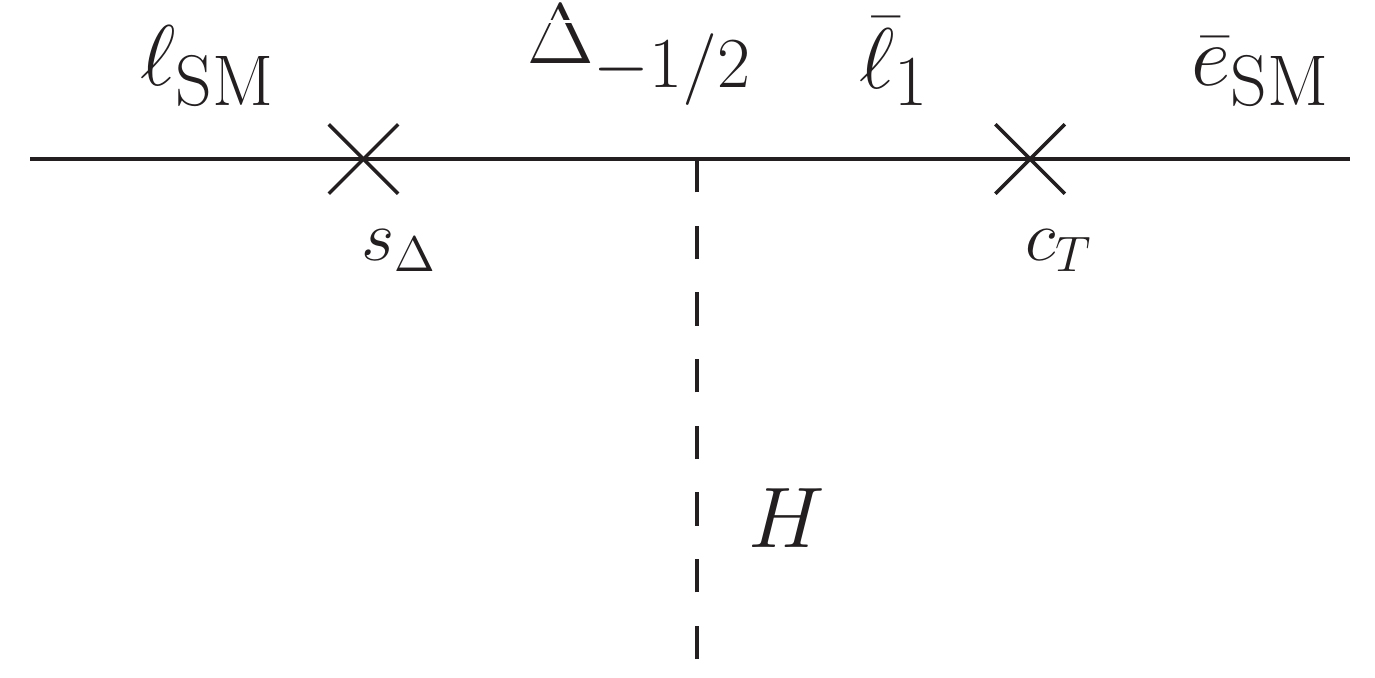} \\
\vspace{0.3cm}
\includegraphics[width=0.4\linewidth]{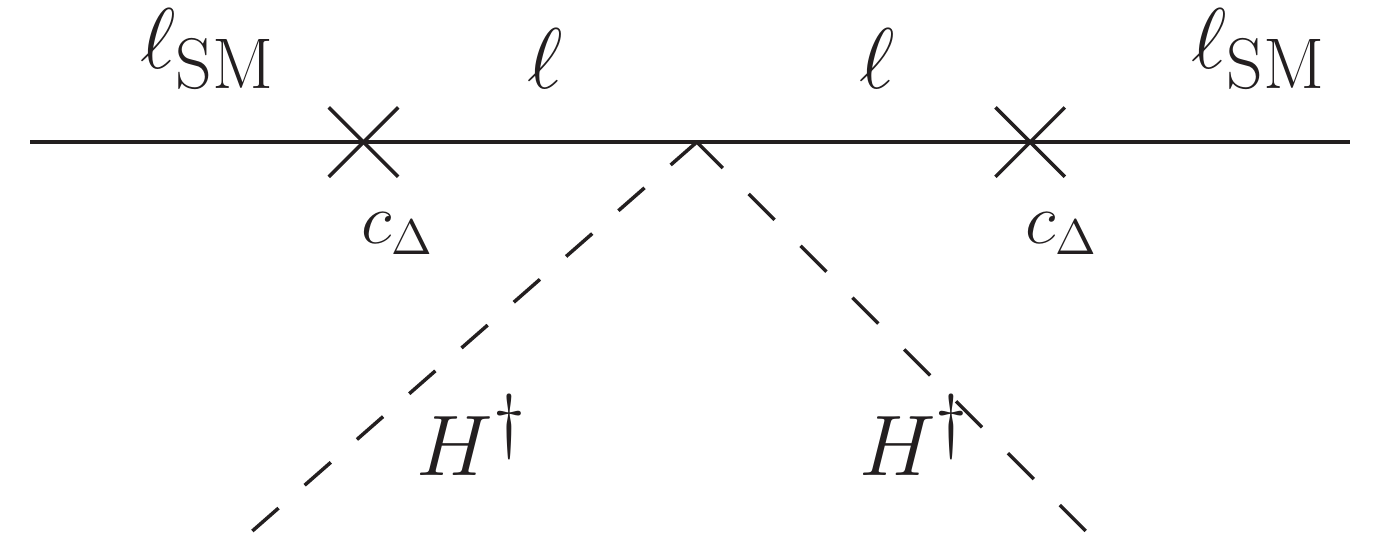}
\caption{The generation of SM fermion masses. The mixing angles are defined so that they vanish in the limit $m_X \gg x_X v'$, $s_X = x_X v' / \sqrt{m_X^2 + x_X^2 {v'}^2}$.}
\label{fig:yukawa}
\end{figure}

Here and below we neglect flavor mixing, which can be straightforwardly taken into account, so that $x_Q$ and $m_Q$ are real parameters referring to a single generation.
We show how the SM up-type yukawa coupling arises in the upper most panel of Fig.~\ref{fig:yukawa}.
Because of the non-zero $\vev{H'}$, $Q_{1/6}$ and $q$ mix with each other. The mixings in Fig.~\ref{fig:yukawa} are given by
\begin{align}
s_X \equiv {\rm sin} \, \theta_X = \frac{x_X v'}{\sqrt{m_X^2 + x_X^2 {v'}^2}}.
\end{align}
A linear combination of $Q_{1/6}$ and $q$ obtains a mass $\sqrt{m_Q^2 + x_Q^2 {v'}^2}$, paired with $\bar{Q}_{-1/6}$. The orthogonal linear combination of $Q_{1/6}$ and $q$ becomes a doublet quark of the SM acquiring a yukawa coupling to $\bar{q}_{-2/3}$, a right-handed up-type quark, of
\begin{align}
y_u = x_Q s_Q = \frac{x_Q^2 {v'}}{ \sqrt{ m_Q^2 + x_Q^2 {v'}^2}}.
\end{align}
Except for the top yukawa coupling, we expect $m_Q \gg x_Q v'$ to be a good approximation, so that $y_u \simeq x_Q^2 v' / m_Q $ for the up and charm quarks.
$Q_{-5/6}$ and $\bar{Q}_{5/6}$ obtain a mass $m_Q$.

When the $X$ states arise from ${\bf 45}$, we have
\begin{align}
\label{eq:uyukawa45}
{\cal L }_{45_{QU}} =  \bar{q} \, x_Q Q \,  H^\dag   +  q \, x_Q \bar{Q} \, {H'}^\dag   + m_Q Q \bar{Q} +   q \, x_U \bar{U} \,  H^\dag  +   \bar{q} \, x_UU \,  {H'}^\dag  + m_U U \bar{U}.
\end{align}
The fate of $Q\bar{Q}$ and $q$ are the same as for ${\bf 54}$. A linear combination of $\bar{q}_{-2/3}$ and $\bar{U}$ pairs with $U$ and obtains a mass $\sqrt{m_U^2 + x_U^2 {v'}^2}$. The orthogonal combination becomes a $\bar{u}$ of the SM, so that the corresponding up-type yukawa coupling is
\begin{align}
y_u = x_Q s_Q c_U + x_U c_Q s_U = \frac{(x_Q^2 m_U + x_U^2 m_Q) v'}{\sqrt{m_U^2 + x_U^2 {v'}^2} \sqrt{m_Q^2 + x_Q^2 {v'}^2}}.
\end{align}
Small up and charm yukawa couplings are explained by $m_{Q,U} \gg x_{Q,U} v'$ or $m_{Q,U} \ll x_{Q,U} v'$.

\subsection{Down-type quark yukawa coupling}
The $X$ states for down-type quark yukawa couplings are in ${\bf 10}$ of $SO(10)$, as larger representations result in non-perturbative gauge couplings.  Yukawa couplings arise from
\begin{align}
{\cal L }_{10_D} =   q \, x_D\bar{D} \, H  +  \bar{q} \, x_D D {H'}   + m_D D \bar{D}.
\end{align}
A linear combination of $\bar{q}_{1/3}$ and $\bar{D}$ obtains a mass $\sqrt{m_D^2 + x_D^2 {v'}^2}$, paired with $D$. The orthogonal linear combination is the SM right-handed down quark.
The SM down-type yukawa coupling is
\begin{align}
\label{eq:dyukawa}
y_d = x_D c_Q s_D = \frac{x_D^2 {v'}}{ \sqrt{ m_D^2 + x_D^2 {v'}^2}}\frac{m_Q}{ \sqrt{ m_Q^2 + x_Q^2 {v'}^2}}.
\end{align}

\subsection{Charged lepton yukawa couplings}
The $X$ states for charged lepton yukawa couplings are also in ${\bf 10}$ of $SO(10)$, and the yukawa couplings arise from 
\begin{align}
{\cal L}_{10_\Delta} = \bar{\ell} \, x_\Delta \Delta \, H  + \ell \, x_\Delta \Delta \, H'  + \frac{1}{2} m_\Delta \Delta^2.
\end{align}
A linear combination of $\Delta_{-1/2}$ and $\ell$ obtains a mass $\sqrt{m_\Delta^2 + y_\Delta^2 {v'}^2}$, paired with $\Delta_{1/2}$. The orthogonal linear combination is the SM lepton doublet.

The SM charged lepton yukawa couplings depend on whether the $X$ states for the up-type quark is ${\bf 54}$ or ${\bf 45}$. If it is ${\bf 54}$, $\bar{\ell}_{1}$ is the SM right-handed charged lepton. If it is ${\bf 45}$, we need to take into account the following interaction,
\begin{align}
\label{eq:45lepton}
{\cal L }_{45_T} =  \ell \, x_T T_L \, H^\dag   +  \bar{\ell}\, x_TT_R \, {H'}^\dag  + \frac{1}{2}m_T T_L^2  + \frac{1}{2}m_T T_R^2.
\end{align}
A linear combination of $\bar{\ell}_1$ and $T_{R,1}$ obtains a mass $\sqrt{m_T^2 + y_T^2 {v'}^2}$, paired with $T_{R,-1}$. The orthogonal linear combination is the SM right-handed charged lepton. The SM charged lepton yukawa coupling is
\begin{align}
\label{eq:eyukawa}
y_e =
x_\Delta s_\Delta \times
\begin{cases}
1  \\
c_T  
\end{cases}
=
\frac{x_\Delta^2 v' }{ \sqrt{ m_\Delta^2 + x_\Delta^2 {v'}^2}} \times
\begin{cases}
1 & : X_{54} \\
\frac{m_T}{ \sqrt{ m_T^2 + x_T^2 {v'}^2}} & : X_{45}
\end{cases}.
\end{align}

\subsection{Neutrino masses and mixing}
If the $X$ states of up-type quarks are ${\bf 54}$, neutrino masses may arise from the interactions and mass term,
\begin{align}
{\cal L}_{54_S} = \ell \, x_S S \, H^\dag   + \bar{\ell} \, x_S S \,{H'}^\dag  + \frac{1}{2} m_S S^2.
\label{eq:54S}
\end{align}
In the 422 theory $m_S = m_U$ and $x_S = \sqrt{3}/2 \, x_U $.
At tree-level, only one linear combination of $\nu$ and $N$, which is predominantly $N$, obtains a mass, and the SM neutrino remains massless. However, since lepton and chiral symmetries are broken by $m_S$, there is no symmetry forbidding a SM neutrino mass, which should arise from quantum corrections. Taking into account mixing between $\ell$ and $\Delta_{-1/2}$, the neutrino mass is
\begin{align}
m_\nu \sim \frac{1}{16\pi^2} \frac{x_S^2 v^2}{ m_S} c_\Delta^2 =  \frac{1}{16\pi^2} \frac{x_S^2 v^2}{ m_S} \frac{m_\Delta^2}{m_\Delta^2 + x_\Delta^2 {v'}^2 }~.
\end{align}
If the $X$ states of up-type quarks are ${\bf 45}$, the neutrino mass arises from Eq.~(\ref{eq:45lepton}),
\begin{align}
m_\nu =\frac{x_T^2 v^2}{ m_T} \frac{m_\Delta^2}{m_\Delta^2 + x_\Delta^2 {v'}^2 }~.
\end{align}

Next we consider aspects of flavor mixing.  Although the same $SO(10)$ states, $X_{45,54}$, contribute to both up-type quark and neutrino masses, the lack of mass hierarchies and large mixing angles of neutrinos compared with up-type quarks can be understood. Assuming $m_{\Delta} \ll x_{\Delta} v'$ only for the third generation, the neutrino mass matrix is given by
\begin{align}
m_\nu \sim \left( \frac{1}{16\pi^2} \right)_{54} \;\; \frac{v^2}{v'}
\begin{pmatrix}
1 & & \\
 & 1 & \\
 & & c_{\Delta_3}
\end{pmatrix}
y_u 
\begin{pmatrix}
1 & & \\
 & 1 & \\
 & & c_{\Delta_3}
\end{pmatrix},
\hspace{0.25in} ~~ c_{\Delta_3} =  \frac{m_{\Delta_3}}{\sqrt{m_{\Delta_3}^2 + x_{\Delta_3}^2 {v'}^2 }} \ll1
\label{eq:mnu}
\end{align}
where the factor of $1/16\pi^2$ applies only for $X_{54}$. With $c_{\Delta_3} = O(10^{-2})$, the neutrino mass matrix is not near-diagonal nor hierarchical, except for the $(1,1)$ component.  Thus, in Higgs Parity Unification we are able to derive an order-of-magnitude mass relation
\begin{align}
m_{\nu_{2,3}} \sim  \left( \frac{1}{16\pi^2} \right)_{54} \;\; \frac{v}{v'} \; m_c
\end{align}
which is successful since the Higgs mass and coupling unification require $v' = 10^{10\mathchar`-12}$ GeV. The small up quark mass arises from the $(1,1)$ component of (\ref{eq:mnu}), so that the lightest neutrino is much lighter than the other two neutrinos, goiving a normal hierarchy with
\begin{align}
\frac{m_{\nu_1}} {m_{\nu_{2,3}} } \sim \frac{m_u}{m_c}.
\end{align}

Because of the suppression of the neutrino mass by $c_{\Delta_3}$, for a given mass $m_{S,T}$ the couplings $x_{S,T}$ are larger than expected from the usual see-saw relation $m_\nu \sim x_{S,T}^2 v^2 / m_{S,T}$. We expect that the lepton asymmetry produced by decays of $S$ and $T$ is enhanced, reducing the minimal reheating temperature for successful leptogenesis, whether thermal~\cite{Giudice:2003jh,Buchmuller:2004nz} or non-thermal~\cite{Ibe:2005jf}.

\subsection{A simple $SO(10)$ theory of flavor}
The following renormalizable $SO(10)$ model can economically describe all quark and lepton masses
\begin{align}
\label{eq:SO(10)yukawa}
{\cal L} \; = \; & \psi_{16} \, x_{45} \, X_{45} \, \phi_{16}^\dag + \frac{1}{2} X_{45}(m_{45} + h_{45} \Sigma) X_{45} \nonumber \\
& + \psi_{16} \, x_{10} \, X_{10} \, \phi_{16} + \frac{1}{2} X_{10}(m_{10} + h_{10} \Sigma) X_{10} + {\rm h.c.}~.
\end{align}
Here we introduce three generations of fermions, $\psi_{16}$, $X_{45}$ and $X_{10}$, with generation indices understood. The Higgs $H$ and $H'$ are embedded into $\phi_{16}$. $\Sigma$ is an $SO(10)$ symmetry breaking field, and $x$, $m$ and $h$ are constants.
Since $\Sigma$ does not appear in the above yukawa interactions, this model predicts $x_{Q} = x_U = x_T = 2 / \sqrt{3} x_S$ and $x_D = x_\Delta$ at the $SO(10)$ scale.  However, while mass parameters $m_X$ are not necessarily unified,
we assume that differing 3221 multiplets in the same $X$ representation have masses that are not hierarchically different from each other (e.g.~$m_D \approx m_\Delta$).

Despite the unification of $x_X$, and departures from unification of $m_X$ by only $O(1)$ amounts, the neutrino masses and mixings can be obtained via $c_{\Delta_3} = O(10^{-2})$, as explained in the previous sub-section. This requires that both $s_{\Delta_3} $ and $s_{D_3}$ are very close to unity, and hence $y_b = x_D c_Q$ and $y_\tau = x_\Delta c_T$. Thus $m_b/m_\tau$ differs from that predicted in minimal $SU(5)$ unification schemes by $c_Q /c_T$, which arises from an O(1) difference between $m_Q$ and $m_T$. The ratios $m_d/m_e$ and $m_s/m_\mu$ can also be explained by $m_D/m_\Delta$ ratios at the $SO(10)$ scale that are not far from unity.

The strong CP problem is solved by Left-Right symmetry including space-time parity. Above the $SO(10)$ breaking scale, the symmetry of the theory is $SO(10) \times CP$. This symmetry is broken down to Left-Right symmetry with space-time parity by the VEV of a field that is both Left-Right and CP odd. The CP symmetry requires that $x_{10.45}$ and $m_{10,45}$ are real. When $\Sigma$ is made from an odd number of the Left-Right and CP odd fields, the couplings $h_{10,45}$ are pure-imaginary, explaining the CKM phase.

Ref.~\cite{Hall:2018let} shows that the quantum corrections to the strong CP phase arise at two-loop level. The corrections are shown to be below the current limit from the neutron electric dipole moment under the assumption that the couplings $x$ are $O(1)$ and $m_X$ are above $v'$. This assumption is not valid for the $(3,3)$ components of $x_D \sim y_b$ and $m_D \sim 10^{-2} y_b v'$. As a result, the suppression of the corrections found in~\cite{Hall:2018let} is not guaranteed, and the corrections may be as large as $10^{-6}$. We expect that the corrections are suppressed by an appropriate flavor structure of $x$ and $m_X$, which we leave to future work.

\subsection{The bottom-tau mass ratio and $x_Q$}
\label{sec:btau}

If the top quark mass is generated by $X_{54}$, then the bottom quark but not the tau, receives a suppression from the up-type quark yukawa sector; thus from Eqs.~(\ref{eq:dyukawa}) and (\ref{eq:eyukawa}) $y_b = c_Q x_D s_D$ while $y_\tau = x_\Delta s_\Delta$.
Assuming $x_D = x_\Delta$ at the unification scale, as well as $c_{\Delta,D} \ll 1$ to obtain the neutrino masses and mixing, we find $y_b/y_\tau = c_Q$ at the unified scale, which is renormalized to 
\begin{align}
\frac{y_b}{y_\tau}(m_Z) = 2.1 \frac{m_Q}{ \sqrt{ m_Q^2 + x_Q^2 {v'}^2}}.
\end{align}
To obtain the observed ratio of $1.6$, we need 
\begin{align}
x_Q \simeq 1.5 \, y_t.
\label{eq:xQ54}
\end{align}

If the top quark mass is generated by $X_{45}$, the tau yukawa coupling may be also suppressed by $c_T<1$. To obtain the bottom/tau ratio, $m_T > m_Q$ is required. Unless $m_T \simeq m_Q$, we may neglect the suppression of the tau yukawa coupling. Assuming $x_D = x_\Delta$ at the unification scale as well as $c_{\Delta,D} \ll 1$,
the observed value of $m_b/m_\tau$ then requires
\begin{align}
x_Q \simeq
1.01 \, y_t
\label{eq:xQ45}
\end{align}
where we assume $x_Q \simeq x_U$ and $m_Q \simeq m_U$.

The values of $x_Q$ from (\ref{eq:xQ54}) and (\ref{eq:xQ45}) are used to evaluate the top quark threshold correction to $\lambda_{\rm SM} (v')$ in Sec.~\ref{sec:scale}.

\section{Prediction for the Scale of Parity Breaking }
\label{sec:scale}
In Sec.~\ref{sec:HParity}, we showed that the SM Higgs quartic coupling essentially vanishes at tree level at the Higgs Parity symmetry breaking scale, $v'$. In this section, we compute threshold corrections to the Higgs quartic coupling and derive $v'$ in terms of SM parameters.

\subsection{Threshold corrections to the SM quartic coupling}

The tree-level scalar potential is
\begin{align}
V_{\rm tree} = & \lambda \left( |H|^2 + |H'|^2 \right)^2 + \lambda'|H|^2 |H'|^2 - m^2 (|H|^2 + |H'|^2).
\end{align}
After taking into account quantum corrections, the coupling $\lambda'(v')$ becomes non-zero.

\let\oldaddcontentsline\addcontentsline
\renewcommand{\addcontentsline}[3]{}

\subsubsection{Threshold correction from charged gauge bosons}
The one-loop Coleman-Weinberg (CW) potential from $W$ and $W'$ bosons is 
\begin{align}
V_{1-{\rm loop}} =  c |H|^4 \, {\rm ln} \frac{|H|}{M} + c |H'|^4 \, {\rm ln} \frac{|H'|}{M},~~c \equiv &  \frac{3}{64\pi^2}g^4,
\end{align}
where $M$ is an arbitrary scale. A change of $M$ can be absorbed by a change of $\lambda$. The vev of $H'$ satisfies
\begin{align}
m^2 =2 \lambda {v'}^2 \left(1 + \frac{c}{4 \lambda} + \frac{c}{2\lambda} {\rm ln} \frac{{v'}^2}{M^2}  \right)
\end{align}
After integrating out $H'$, the potential for $H$, to leading order in $c$ and $\lambda'$, is given by
\begin{align}
V(H) \simeq {v'}^2\left(\lambda' - \frac{c}{2} - c \, {\rm ln} \frac{{v'}^2}{M^2}  \right) \, |H|^2 + \left( - \lambda' + \frac{3}{4} c + c \, {\rm ln} \frac{{v'}^2}{M^2}  + c \, {\rm ln} \frac{|H|}{v'}  \right)\, |H|^4.
\end{align}
To obtain the electroweak scale much smaller than $v'$ requires $\lambda' \simeq c /2 + 2c \, {\rm ln}(v'/M)$, giving
\begin{align}
V(H) / |H|^4  \simeq \frac{c}{4} \, ( 1 + 4 \, {\rm ln} \frac{|H|}{v'} ).
\end{align}

We match this potential to the one-loop CW potential of the SM from the $W$ boson,
\begin{align}
V_{\rm SM}(H)/ |H|^4  = \, &\lambda_{\rm SM}(\mu)
+ \frac{3}{128\pi^2} g^4 \left( {\rm ln}\frac{ g^2 |H|^2 / 2}{ \mu^2} - \frac{3}{2}  \right), \nonumber
\end{align}
where we take the $\overline{\rm MS}$ scheme. 
By matching $V_{\rm SM}(H)$ to  $V(H)$ with $\mu = v'$, we obtain
\begin{align}
\lambda_{{\rm SM},W}(v') \simeq  \frac{3}{64\pi^2} g^4 \, {\rm ln} \frac{e}{ g/ \sqrt{2}}.
\end{align}
To suppress higher order corrections, the coupling $g$ should be evaluated around  $v'$.

\subsubsection{Threshold correction from neutral gauge bosons}
The threshold correction from $Z$ and $Z'$ bosons can be estimated in a similar manner.  After integrating out $H'$ and fine-tuning for the electroweak scale, the Higgs potential is
\begin{align}
V(H) / |H|^4  \simeq \frac{3 (g^2 + {g'}^2)^2}{512 \pi^2} \left( 1 + 4 \, {\rm ln} \frac{|H|}{v'} - 2 {\rm ln} \frac{g^4}{ g^4 - {g'}^4} \right).
\end{align}
The one-loop CW potential of the SM from the $Z$ boson is 
\begin{align}
V_{\rm SM}(H)/ |H|^4  = & \, \lambda_{\rm SM}(\mu)+ \frac{3}{256\pi^2} (g^2 + {g'}^2)^2 \left( {\rm ln}\frac{ (g^2 + {g'}^2) |H|^2 / 2}{ \mu^2} - \frac{3}{2}  \right).
\end{align}
Matching these results at $v'$ gives the threshold correction 
\begin{align}
\lambda_{{\rm SM},Z}(v') \simeq \frac{3}{256\pi^2} (g^2 + {g'}^2)^2  \left( {\rm ln} \frac{e^2}{(g^2 + {g'}^2) / 2 }  - {\rm ln} \frac{g^4}{ g^4 - {g'}^4}   \right).
\end{align}

\subsubsection{Threshold correction from top quarks}
The threshold correction from top quarks is model-dependent. Let us first consider the case where the $X$ state for the top quark is ${\bf 54}$, where the relevant interaction is shown in Eq.~(\ref{eq:uyukawa54}).
The mass squareds of the six mass eigenstates are
\begin{align}
& 0, m_Q^2, m_Q^2 + x_Q^2 |H|^2, m_Q^2 + x_Q^2 |H'|^2 ,  \nonumber \\
& \frac{1}{2} \left(   m_Q^2 + x_Q^2 |H|^2 + x_Q^2 |H'|^2 \pm \sqrt{\left(  m_Q^2 + x_Q^2 |H|^2 + x_Q^2 |H'|^2    \right)^2  + 4 x_Q^4 |H|^2 |H'|^2} \right).
\end{align}
With a similar computation to the gauge contribution, we find that
\begin{align}
V(H) / |H|^4  \simeq - \frac{3}{32\pi^2}  y_t^4\left( 1 + 4 \, {\rm ln} \frac{|H|}{v'} + 4  f_{54}\left( \frac{x_Q}{y_t} \right) \right) ,\\
f_{54}(r) = r^4 - {\rm ln} \, r^2 + (r^6 - \frac{r^4}{2}) {\rm ln}( 1- \frac{1}{r^2}).
\end{align}
The one-loop CW potential of the SM from the top quark is 
\begin{align}
V_{\rm SM}(H)/ |H|^4  = & \, \lambda_{\rm SM}(\mu)
- \frac{3}{16\pi^2} y_t^4\left( {\rm ln}\frac{y_t^2 |H|^2}{ \mu^2} - \frac{3}{2}  \right),
\end{align}
and matching these potentials yields the threshold correction 
\begin{align}
\lambda_{{\rm SM,t54}}(v') \simeq - \frac{3}{8\pi^2} y_t^4 \left(  {\rm ln} \frac{e}{y_t}  + f_{54}\left( \frac{x_Q}{y_t} \right)    \right).
\end{align}
Note that the threshold correction logarithmically diverges as $x_Q \rightarrow y_t$ i.e.~$m_Q \ll x_Q  v'$ because, for $m_Q \ll x_Q  v'$, there is an additional particle below the scale $v'$ coupling to the SM Higgs.

We next consider the case where the $X$ state for the top quark is ${\bf 45}$, where the relevant interactions are shown in Eq.~(\ref{eq:uyukawa45}).
We only consider the case $x_U \simeq x_Q$ and $m_{U} \simeq m_Q $, giving the top yukawa coupling
\begin{align}
y_t = \frac{2 x_Q^2 m_Q v'}{m_Q^2 + x_Q^2 {v'}^2}
\end{align}
which can be solved for
\begin{align}
m_Q \simeq\frac{x_Q^2 v'}{y_t}\left( 1 \pm  \sqrt{ 1 - \frac{y_t^2}{x_Q^2}   } \right).
\end{align}
We find the threshold correction
\begin{align}
\lambda_{{\rm SM,t45\pm}}(v') \simeq - \frac{3}{8\pi^2} y_t^4 \left(  {\rm ln} \frac{e}{y_t}  + f_{45\pm}\left( \frac{x_Q}{y_t} \right)    \right),
\end{align}
where the functions $f_{45\pm}$ are given by
\begin{align}
f_{45\pm}(r) = \frac{5}{12} + r^4 + r^4 \left(-  \frac{1}{2} + 2 r^2 \pm 2 r \sqrt{r^2 -1 } \right) {\rm ln} \frac{r \pm \sqrt{r^2 -1 }}{2 r} - \frac{1}{2} {\rm ln}\left(  2 r^3 (r \pm \sqrt{r^2 -1 } ) \right).
\end{align}
The function $f_{45} (r)$ nearly vanishes around $r=1$.

\subsubsection{Threshold correction from other fermions}
The threshold correction from other charged fermions are expected to be negligibly small because the corresponding $m_X \gg v'$ or $x \ll 1$.
An exceptional case arises if the up/charm yukawas, arise from $X_{45}$, and $y_{u,c} \ll 1$ follows from $m_X \ll x v'$ while $x= O(1)$. We do not consider such a case.

The threshold correction from the neutrino that is in the same ${\bf 16}$ as the top quark can be large since $x\sim 1$ and $m_X \sim v'$.
If the $X$ state for the top quark is ${\bf 54}$, the yukawa coupling of $H$ and $H'$ to $\ell$ and $\bar{\ell}$ are of the form (\ref{eq:54S}),
which is $SU(4)$ symmetric. The threshold correction to $\lambda_{\rm SM}(v')$ vanishes at one-loop level. 
If the $X$ state for the top quark is ${\bf 45}$, the yukawa coupling is of the form
\begin{align}
x_T H^\dag \ell T_L + \frac{1}{2}m_T T_L^2  + x_T {H'}^\dag \bar{\ell} T_R + \frac{1}{2}m_T T_R^2
\end{align}
giving the threshold correction
\begin{align}
\lambda_{{\rm SM},\nu45}(v') = &  \frac{x_T^4}{128 \pi^2} f_{45\nu}(\frac{m_T}{ x_T v'}), \nonumber \\
f_{45\nu} (r) = &-\frac{4 \left(6 r^2+11\right)}{r^2+2} - 2 \left(4 r^2+1\right) {\rm ln} \, 2- 2 \left(12 r^2+5\right) {\rm ln} \, r^2  \nonumber \\
 & + 8 \left(2 r^2+1\right) {\rm ln}\left(r^2+1\right) +\frac{r \left(r^2+1\right) \left(4 r^2+9\right) {\rm ln} \frac{r^2+\sqrt{r^2+2} r+1}{r^2-\sqrt{r^2+2} r+1}}{\left(r^2+2\right)^{3/2}}.
\end{align}
We find this correction to be negligible unless $m_T \ll x_T v'$, which requires $m_T \ll m_{Q,U}$ and is disfavored from the bottom-tau ratio.

\subsubsection{Threshold correction from colored Higgses in the 422 theory}
In the 422 theory, colored Higgs have masses around the scale $v'$, and contribute to the threshold correction to $\lambda_{\rm SM}(v')$. We denote the $({\bf 4}, {\bf 2}, {\bf 1})$ and $({\bf \bar{4}}, {\bf 1}, {\bf 2})$  Higgses as $\Phi^a_\alpha$ and $\Phi'_{a,\alpha'}$, respectively. Here $a$, $\alpha$, $\alpha'$ are the $SU(4)$, $SU(2)_L$, $SU(2)_R$ indices, respectively.
The $SU(4)\times SU(2)_L\times SU(2)_R\times P_{LR}$ invariant potential is given in general by
\begin{align}
V = & - m^2 \left( |\Phi|^2 + |\Phi'|^2 \right) + \frac{\lambda}{2} \left( |\Phi|^4 + |\Phi'|^4    \right) + y |\Phi|^2 |\Phi'|^2 + \frac{k}{2} \left(  \Phi^a_\alpha \Phi^{b \alpha}  \Phi_{a\beta}^{*} \Phi_{b}^{*\beta} + \Phi^{'a}_{\alpha'} \Phi^{'b \alpha'}  \Phi_{a\beta'}^{'*} \Phi_{b}^{'*\beta'}  \right) \nonumber \\
 & + \left( \frac{l}{2} \Phi^a_\alpha \Phi^{b \alpha} \Phi_a^{'\beta'} \Phi'_{b \beta'} + {\rm h.c.} \right) + g \Phi^a_\alpha \Phi^{*\alpha}_b \Phi'_{a\beta'} \Phi^{'*\beta'},
\end{align}
where $|\Phi|^2 \equiv \Phi^{a\alpha} \Phi^*_{a \alpha}$. The threshold correction is given by~\cite{Hall:2018let}
\begin{align}
\lambda_{\rm SM}(v') = & \frac{1}{64 \pi^2} |l|^2 \; f_{c} \left( \frac{g}{|l|} , \frac{k}{|l|} \right), \\
f_c(x,y) = & \frac{(1 - (x-y)^2)^2}{6 (x-y)^3}  \left( 2 \left(x-y\right) + \left(x +y \right) {\rm ln} \frac{y}{x}  \right).
\end{align}
The function $f_c$ is always negative and is typically $O(1)$. As long as $|\ell|$, $g$, $k$ are less than $1$, this contribution is subdominant. If $|\ell|$, $g$, $k$ are larger than unity, which leads to strongly coupled Higgses at higher energy scales, this contribution can be large and predicts larger top quark masses. We assume weakly coupled Higgs bosons and neglect the threshold correction from the colored Higgses.

\let\addcontentsline\oldaddcontentsline

\subsection{Top quark mass, QCD coupling and the Higgs Parity breaking scale}

\begin{figure}[t]
\includegraphics[width=0.45 \linewidth]{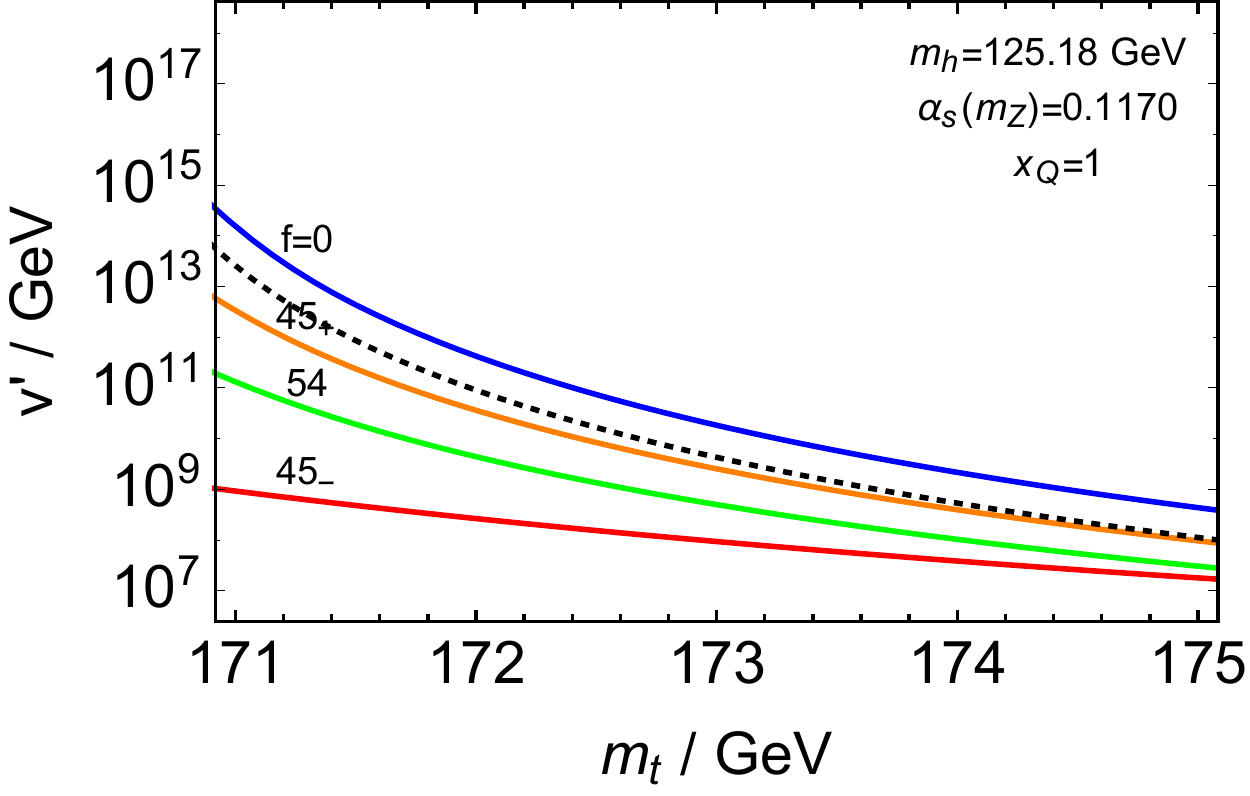}
\includegraphics[width=0.45 \linewidth]{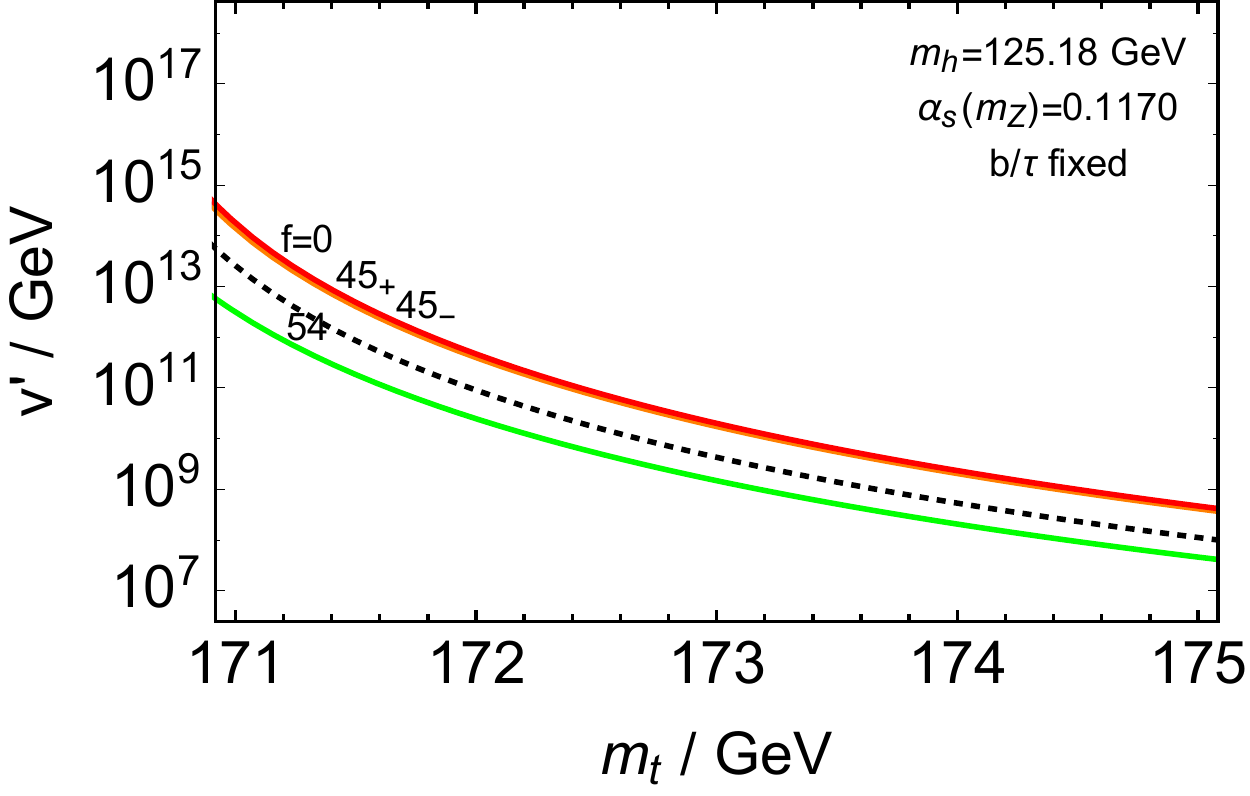}
\includegraphics[width=0.45 \linewidth]{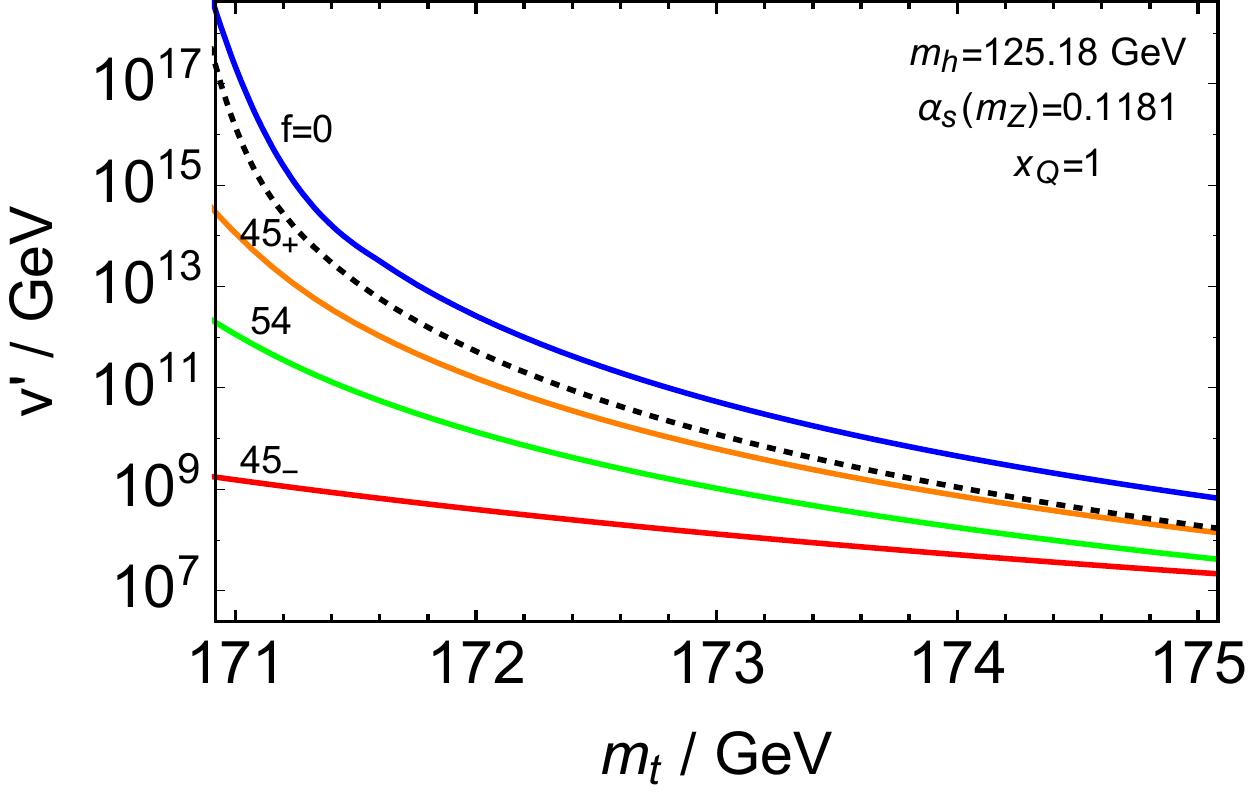}
\includegraphics[width=0.45 \linewidth]{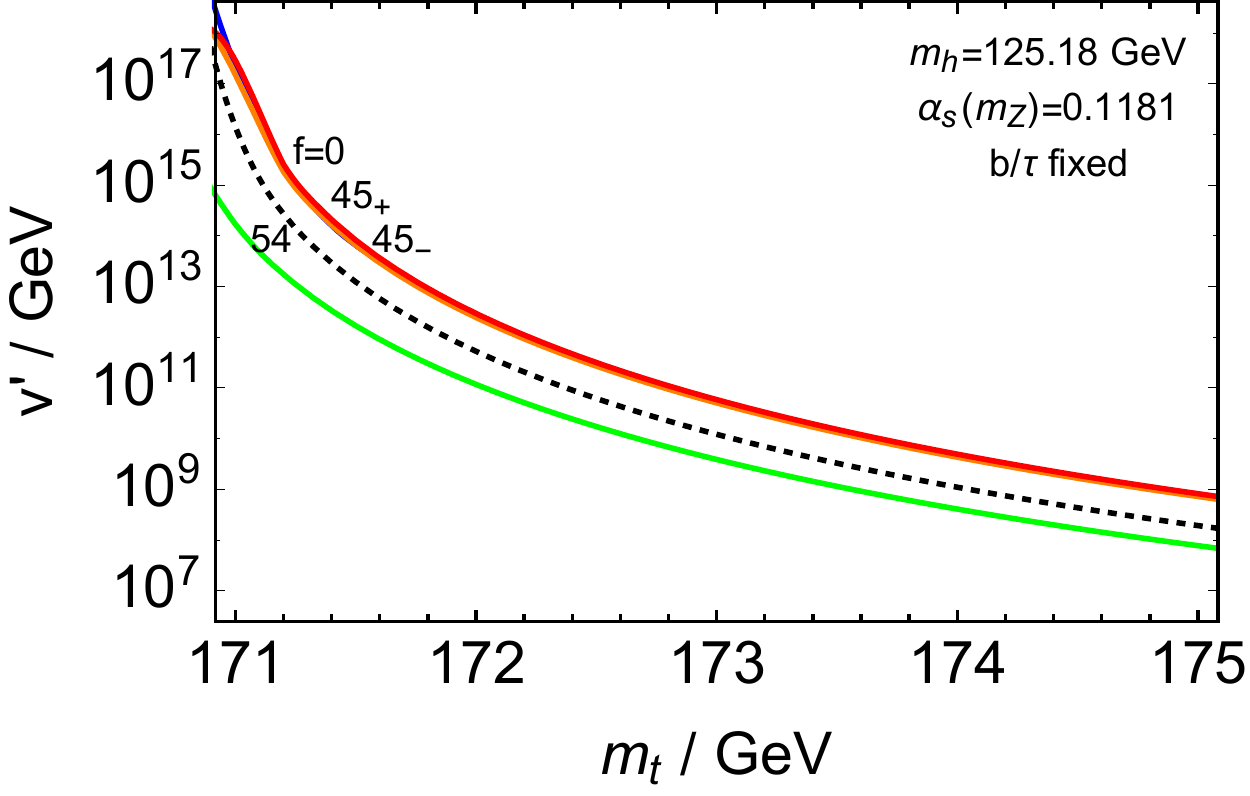}
\includegraphics[width=0.45 \linewidth]{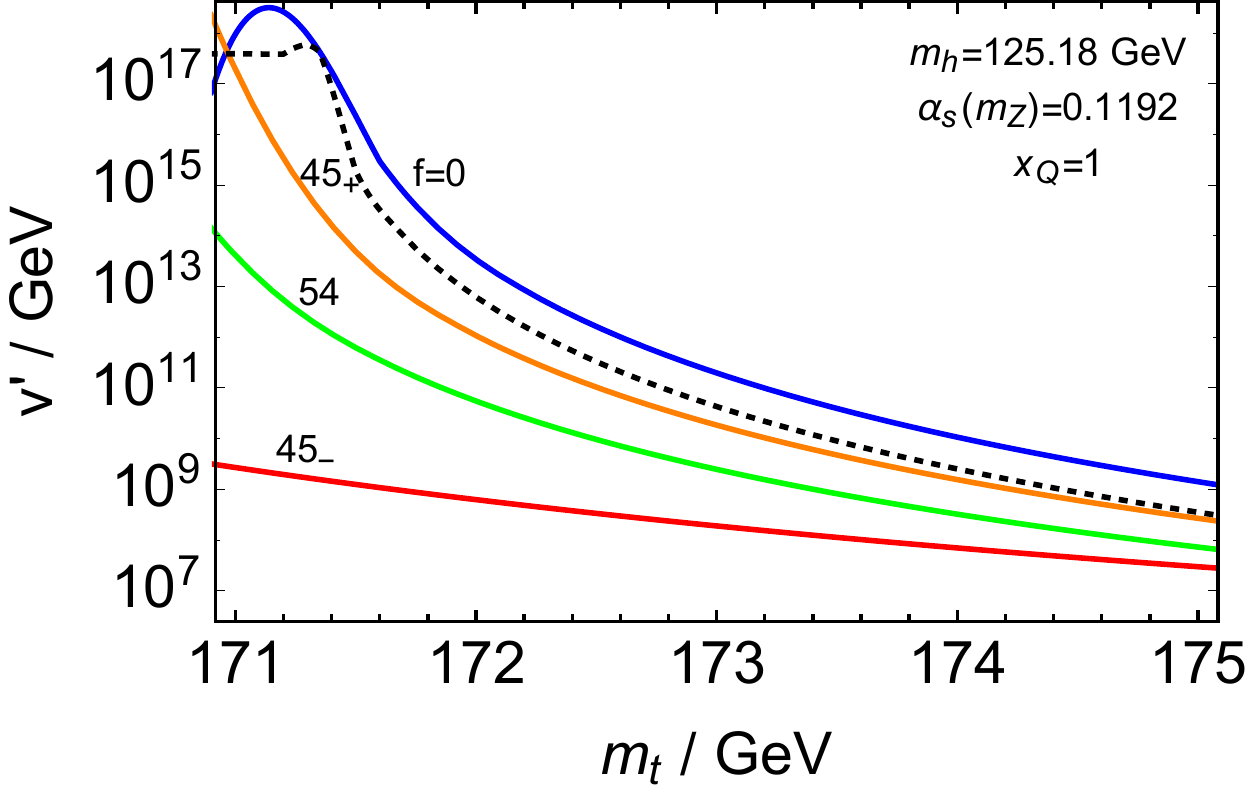}
\includegraphics[width=0.45 \linewidth]{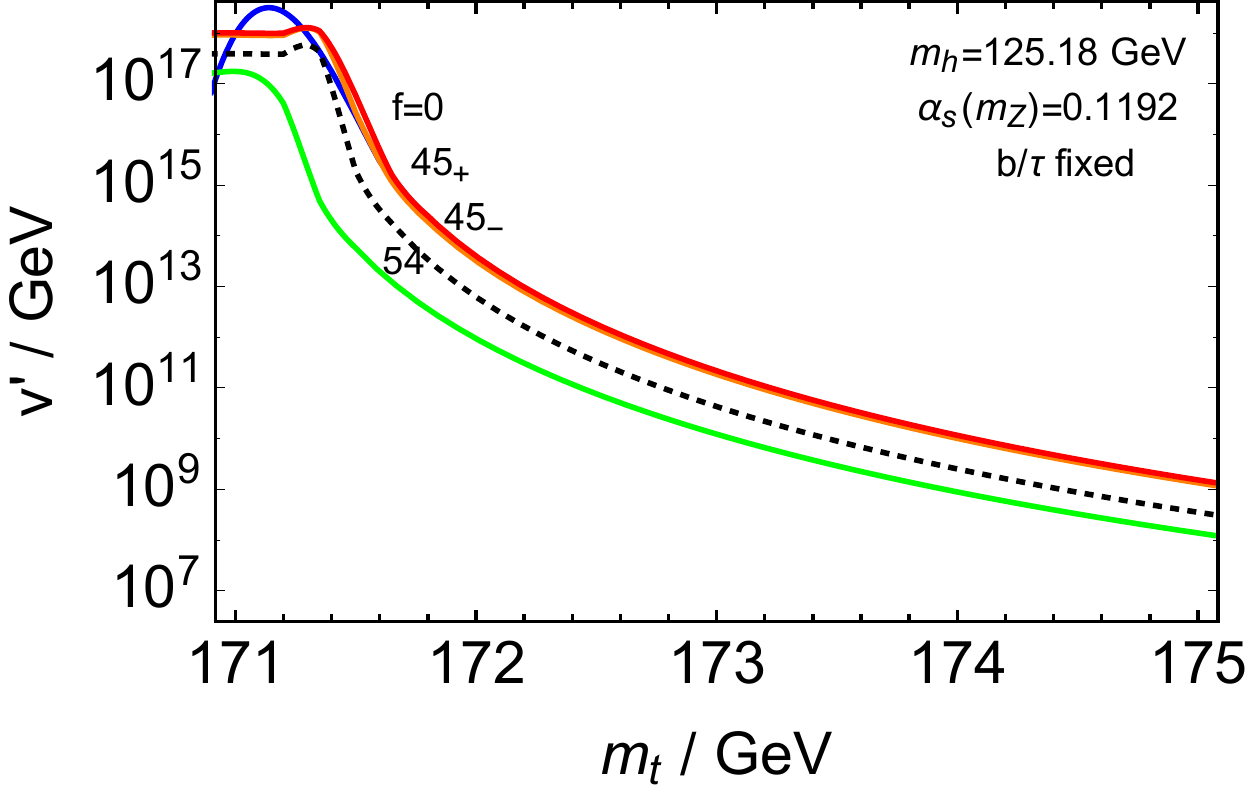}
\caption{The prediction of $v'$, from running of the SM quartic, as a function of $m_t$ with the three rows showing different values of $\alpha_s$. Dotted lines assume $\lambda_{\rm SM}(v')=0$. 
Solid lines show that the dependence of $v'$ on the $X$ states of the top quark is large for $x_Q=1$ (left panels), but is reduced when $m_b/m_\tau$ is imposed (right panels), which also significantly raises the $v'$ prediction.}
\label{fig:vpmt}
\end{figure}

Let us first clarify the top quark mass we use in this paper. We use the pole top quark mass $m_t$, from which we compute the $\overline{\rm MS}$ top yukawa coupling via~\cite{Buttazzo:2013uya}
\begin{align}
\label{eq:ytopMSbar}
y_t(m_t) = 0.93690 + 0.00556 \left( \frac{m_t}{\rm GeV} - 173.34 \right) - 0.00042 \frac{\alpha_s(m_Z) - 0.1184}{0.0007},
\end{align}
where the NNNLO QCD quantum correction is included. The conversion necessarily involves an uncertainty due to the non-perturbative nature of QCD~\cite{Bigi:1994em,Beneke:1994sw,Beneke:1998ui}. To go beyond the precision limited by this uncertainty, which is expected to be as large as the QCD scale, the top quark mass shown in this paper should be understood as a quantity {\it defined} by Eq.~(\ref{eq:ytopMSbar}).
Also, the pole mass measured using hadronic final states suffers from the uncertainty of soft QCD processes including hadronization~\cite{Skands:2007zg}. We nevertheless show the value suggested in~\cite{Tanabashi:2018oca}, $m_t = 173.0 \pm 0.4$ GeV, as a guide.

We compute the running of the SM Higgs quartic coupling following~\cite{Buttazzo:2013uya}.
In Fig.~\ref{fig:vpmt} we show the prediction for the Higgs Parity breaking scale $v'$ as a function of the top quark mass for various values of the QCD coupling constant and choices of the $X$ states. In the left panels, we take $x_Q=1$.
For a given top quark mass, the prediction for $v'$ is smaller than the one for $f=0$, since $f \lesssim 0$. We find that this is also the case for $X_{45}$ with generic $(x_Q,x_U,m_U,m_Q)$. Thus, for a given $v'$, which is fixed by successful unification, we obtain an upper bound on the top quark mass.

We can make a sharper prediction by assuming bottom-tau unification discussed in Sec.~\ref{sec:btau}. In the right panels, we take the value of $x_Q$ to reproduce the bottom/tau ratio. The predictions for $X_{45}$ are indistinguishable from the one for $f=0$. Here it is assumed that $m_Q = m_U$. We find that this is still the case for $m_U \lesssim m_Q$, while for $m_U>m_Q$ the result approaches that of $X_{54}$. The prediction for $X_{54}$ differs from $f=0$, but not by as much as when $x_Q=1$. For this case of the simplest successful $b/\tau$ result, for a given $v'$ we have two predictions for the top quark mass, which differ from each other by $0.6 \mathchar`-1$ GeV.

\section{Precise Unification and SM parameters}
\label{sec:pGUT}

 In Sec.~\ref{sec:GCU} we used gauge coupling unification to predict the unified mass scale $M_{XY}$ and the Higgs Parity breaking scale $v'$ in terms of unified threshold corrections from gauge particles, $r_{XY}$, and from scalars and fermions, $\Delta$, as shown in Figs.~\ref{fig:3221_vp} and \ref{fig:422_vp}. In Sec.~\ref{sec:scale}, $v'$ was predicted by evolution of the SM quartic, including threshold corrections from this Higgs Parity breaking scale that are sensitive to the top quark coupling $x_Q$, as shown in Fig.~\ref{fig:vpmt}.  By combining these results from Secs.~\ref{sec:GCU} and \ref{sec:scale}, which both depend on whether the $X$ state for the top quark mass is a ${\bf 45}$ or ${\bf 54}$, we are finally ready to discuss the correlation among SM parameters discussed in the introduction. 

\subsection{$SU(3)\times SU(2)\times SU(2) \times U(1)$}

\begin{figure}[t]
\includegraphics[width=0.495 \linewidth]{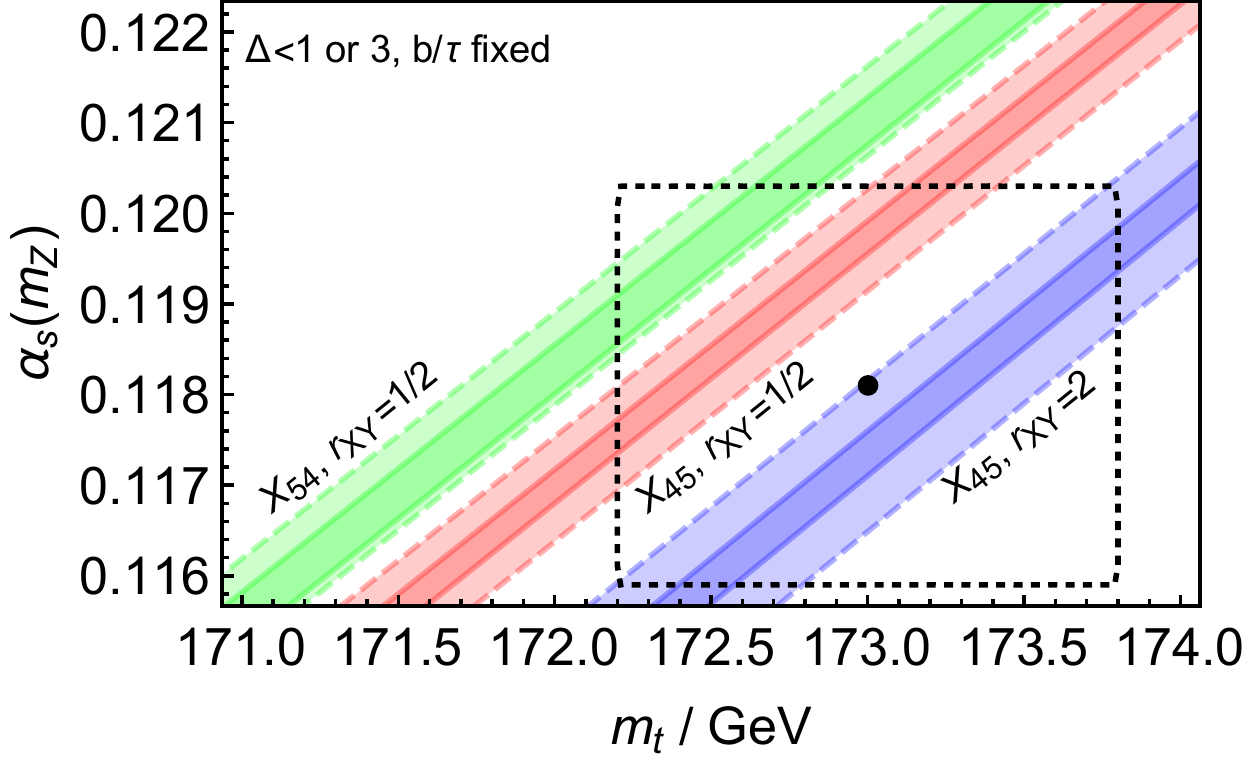}
\includegraphics[width=0.495 \linewidth]{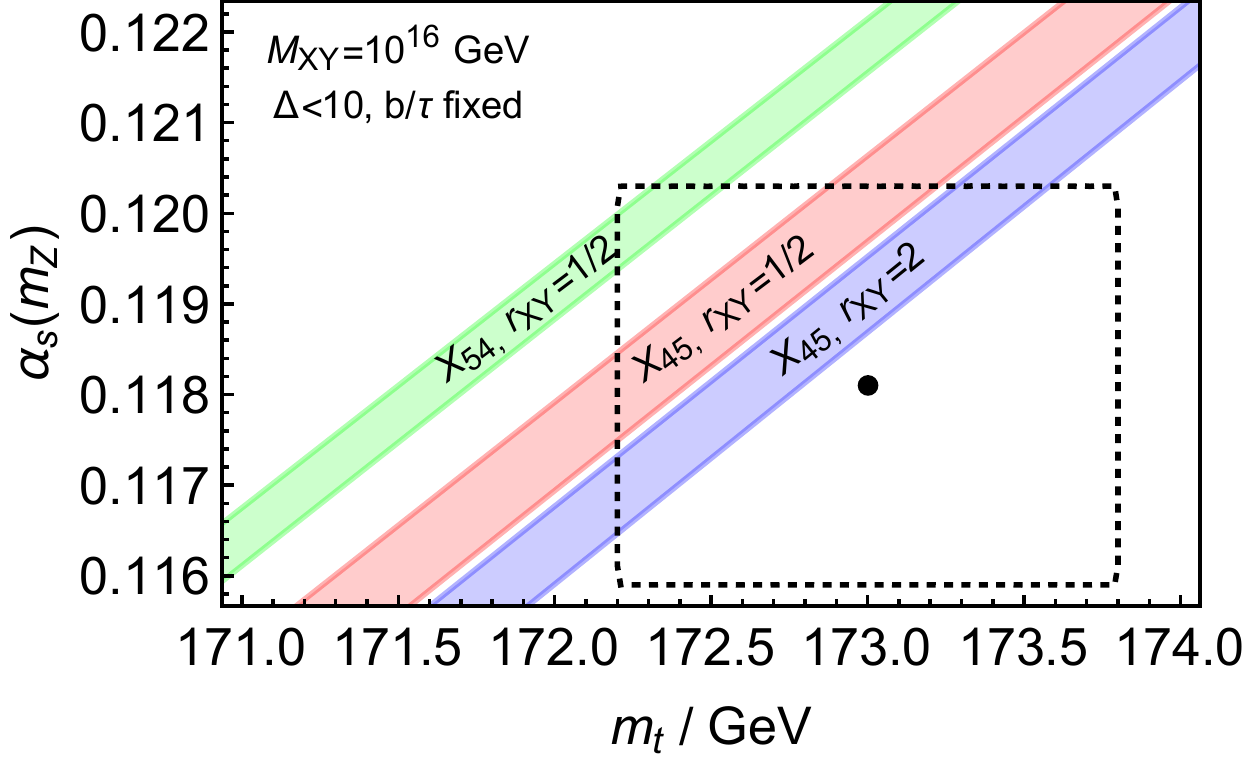}
\caption{Predicted correlation between the top quark mass $m_t$ and the QCD coupling $\alpha_s(m_Z)$ in three 3221 models (left panel); with $M_{XY}$ constrained (right panel). The dot and the dotted box show the central value and the $2\sigma$ range of the observed values, respectively.}
\label{fig:3221_mta3}
\end{figure}

In Fig.~\ref{fig:3221_mta3}, the predicted correlation between $m_t$ and $\alpha_s(m_Z)$ is shown, for $x_Q$ chosen to fix $m_b/m_\tau$.  In the left panel, regions with $\Delta<1$ or 3 are shaded, which is reasonable if the $SO(10)$ Higgses are ${\bf 45}$ or ${\bf 54}$.  For a given theory, $(m_t,\alpha_s)$ is predicted with uncertainties of
\begin{align}
\delta m_t =& \; 0.1 \, {\rm GeV} \times \Delta, \nonumber \\
\delta \alpha_s \simeq& \; 0.0003 \times \Delta.
\end{align}
The $2\sigma$ range of $m_t$ and $\alpha_s(m_Z)$~\cite{Tanabashi:2018oca} is shown by a dotted box.

\begin{figure}[t]
\includegraphics[width=0.5 \linewidth]{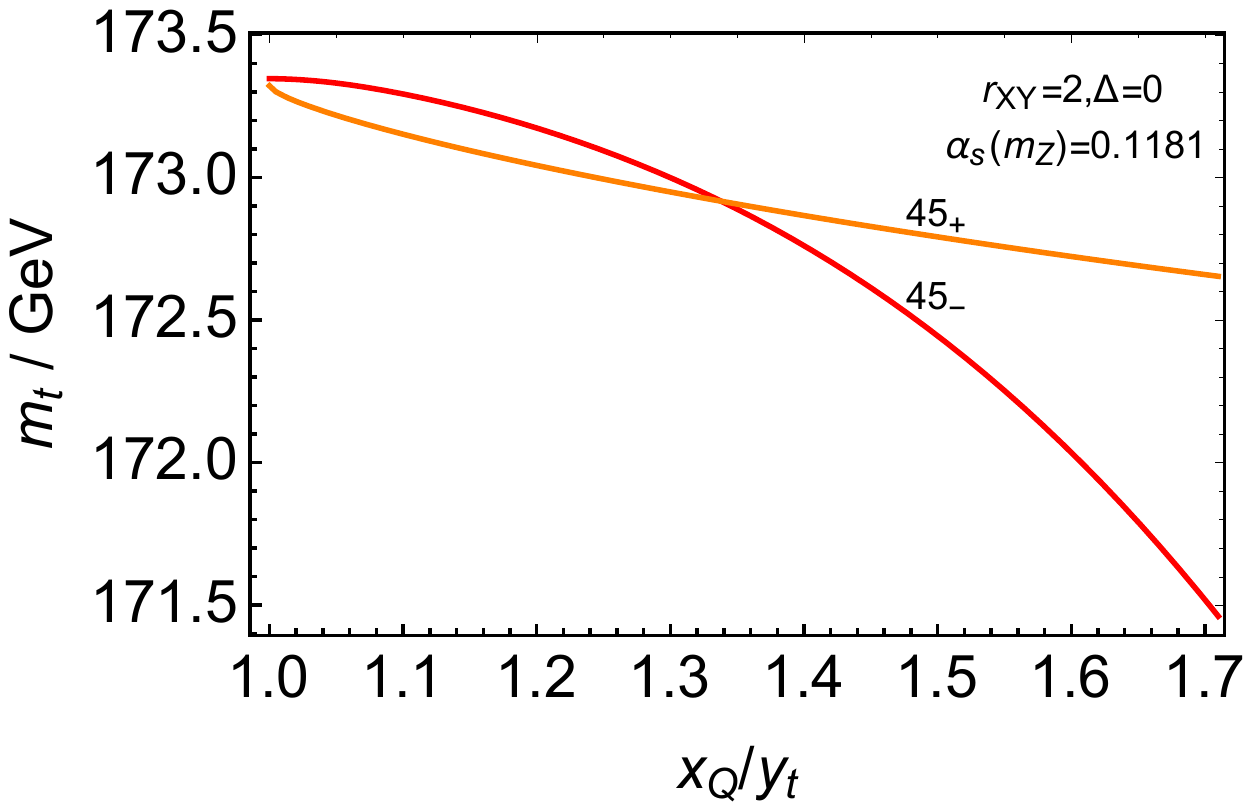}
\caption{The dependence of the predicted top quark mass $m_t$ on $x_Q$ in the 3221 theory.}
\label{fig:xQmt}
\end{figure}

Note that, with $x_Q$ fixed by $m_b/m_\tau$, a top mass from $X_{45}$ gives $f\simeq 0$. Since $f < 0$ for $X_{54 }$ and $r_{XY} \leq 2$ for any breaking to 3221 via $SO(10)$ Higgs of  ${\bf 45, 54, 210}$, the prediction labelled as $X_{45},r_{XY}=2$ can be understood as a model-independent upper bound on the top quark mass. 
For example, if $\alpha_s(m_Z) = 0.1181$, assuming $\Delta <3$, the top quark mass must be below $173.6$ GeV. 
The sensitivity of the prediction on $m_t$ to the value of $x_Q$ is shown in Fig.~\ref{fig:xQmt} for $X_{45}$ and $r_{XY} = 2$. The prediction on $m_{t}$ decreases by $0.3$ GeV if $x_Q$ is larger than the one to fix $m_b/m_\tau$ by more than few 10\%.

The running of the gauge and quartic couplings for the experimental central value of $(\alpha_s, m_t)$  is shown in Fig.~\ref{fig:hpgut}, assuming $X_{45}$, $r_{XY}=2$ and $x_Q$ fixing $m_b/m_\tau$.
The global picture of the correlation shown in Fig.~\ref{fig:global} also assumes the same setup, although the picture looks similar for other choices of the $X$ states, $r_{XY}$ and $x_Q$.

In the right panel of Fig.~\ref{fig:3221_mta3}, we fix the $XY$ gauge boson mass to be $10^{16}$ GeV, which would be suggested if proton decay is observed by Hyper-K. A large value for $\Delta$ is then needed for unification, and the widths of the shaded bands result from requiring $\Delta <10$. The top quark mass must be below $172.8$ GeV for $\alpha_s(m_Z) = 0.1181$.
If proton decay is observed by Hyper-K and the top quark mass is found to be near this bound, we can infer that the bottom-tau ratio is fixed by $x_Q$ and $SO(10)$ symmetry is broken by a ${\bf 45}$ VEV.

\subsection{$SU(4)\times SU(2)\times SU(2)$}

\begin{figure}[t]
\includegraphics[width=0.5 \linewidth]{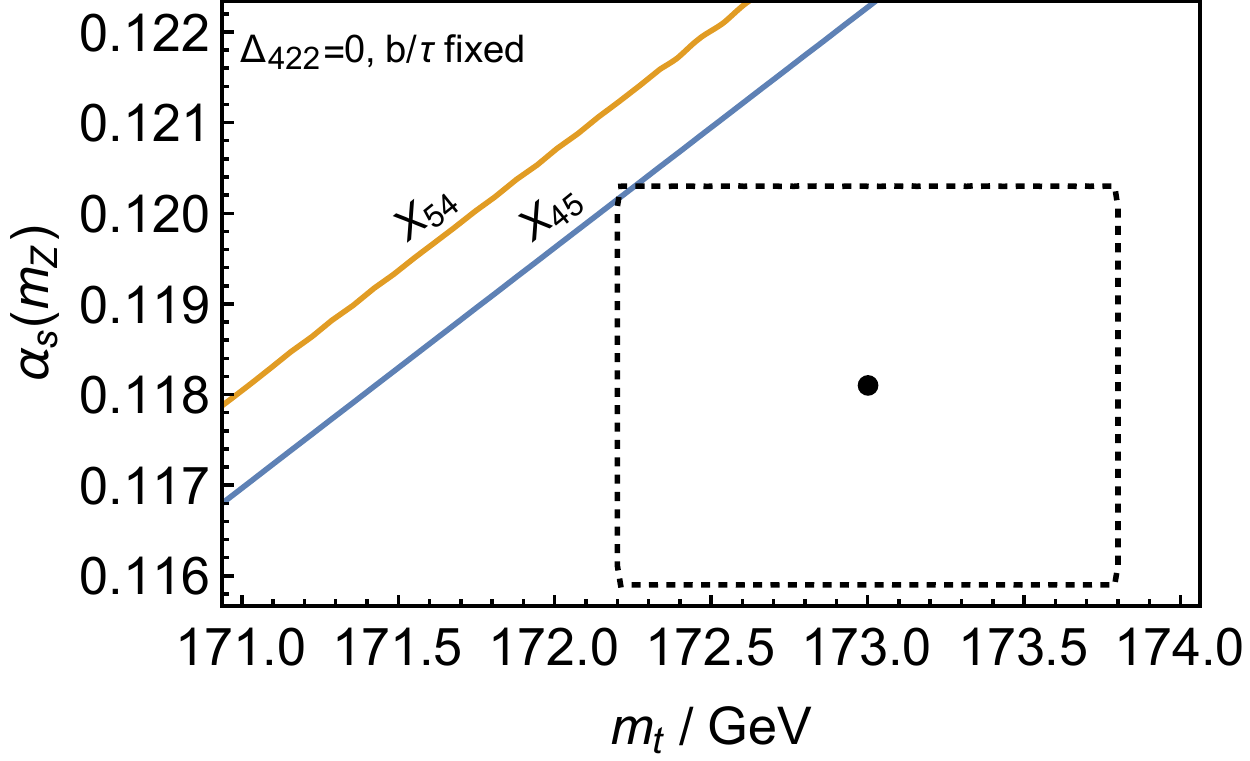}
\caption{Predicted correlation between the top quark mass $m_t$ and the QCD coupling $\alpha_s(m_Z)$ in the 422 theory. The dot and the dotted box show the central value and the $2\sigma$ range of the observed values, respectively.}
\label{fig:422_mta3}
\end{figure}

In the 422 theory, the embedding of the $U(1)_Y$ coupling into the $SU(4)\times SU(2)_R$ couplings is non-trivial.
In the minimal theory, the threshold correction $\Delta_{422}$ only arises from the colored Higgs,
\begin{align}
\Delta_{422} = \frac{2}{5} {\rm ln}\frac{m_{h_Q}}{m_{W'}},
\end{align}
where $m_{h_Q}$ is the mass of the colored Higgs whose gauge quantum number is the same as that of the SM quark doublet. The magnitude of this correction is less than $1$, unless the parameters of the Higgs potential are fine-tuned to make the colored Higgs much lighter than $W'$.
A contribution to $\Delta_{422}$ may also arise from the mass splitting of $X$ states. As long as $x$ and $m_X$ preserve approximate grand unified relations, this contribution is also small. One may wonder whether the hierarchy of $m_X \ll x v'$ leads to a large threshold correction. This is not the case since the VEV of $H'$ breaks $SO(10)$ only to $SU(5)$.

For $\Delta_{422}=0$, $v' \simeq 1.3 \times 10^{14}$ GeV is required. In Fig.~\ref{fig:422_mta3}, the predicted correlation between $m_t$ and $\alpha_s(m_Z)$ is shown, for $x_Q$ chosen to fix $m_b/m_\tau$. Note that, for this choice of $x_Q$, a top mass from $X_{45}$ gives $f\simeq 0$. Since $f\lesssim0$, the prediction labelled as $X_{45}$ can be understood as a model-independent upper bound on the top quark mass. The top quark mass/QCD coupling constant is predicted to be significantly smaller/larger than the central value.

\section{Discussion}

Higgs Parity accounts for a remarkable coincidence: the scale at which the SM quartic coupling vanishes is close to the scale  of Left-Right symmetry breaking required for gauge coupling unification in $SO(10)$, as illustrated in Fig.~\ref{fig:hpgut}.  In this paper we have explored in detail the precision of this coincidence, which we frame in terms of a correlation of the measured values of the top quark mass and the QCD coupling.  

Taking the intermediate gauge symmetry to be 3221, the global picture of this correlation is shown in the right panel of Fig.~\ref{fig:global}, and the fine detail close to the experimental values is shown in the left panel of Fig.~\ref{fig:3221_mta3}.  This correlation is indeed remarkable, and appears at least as precise as the correlation of the QCD coupling with the weak mixing angle in supersymmetric unification.  The constraint on the 3221 breaking scale from gauge coupling unification alone is shown in Fig.~\ref{fig:3221_vp}, and is roughly $v' \simeq (10^{10} \mathchar`-10^{12})$ GeV.  This should be compared with the constraint on $v'$ from running the SM quartic coupling, shown in Fig.~\ref{fig:vpmt}, which is significantly affected by the threshold effect from a coupling $x_Q$ of the top quark sector.  If this parameter is the dominant effect reconciling $m_b/m_\tau$ with unified yukawa couplings, then this constraint on $v'$ is sharpened.  Matching the values of $v'$ from gauge coupling unification and SM quartic running, and allowing typical threshold corrections in simple models of $SO(10)$ breaking, then yields a successful prediction at high precision: $\alpha_s$ to 1\%, or $m_t$ to 0.2\%, as illustrated in the left panel of Fig.~\ref{fig:3221_mta3}.   

The precision may be reduced in more complicated models, or if large $SO(10)$ breaking effects enter the spectrum or couplings of the $X$ states that generate yukawa couplings.  However, as experimental uncertainties on $\alpha_s$ and $m_t$ are reduced, evidence may accumulate for a particular simple version of Higgs Parity unification.  For example, future measurements leading to the blue region of the left panel of Fig.~\ref{fig:3221_mta3} would provide evidence for a simple model with: $SO(10)$ broken via a {\bf 45} to 3221, small unified corrections from scalars and fermions, $X_{45}$ exchange generating the top yukawa coupling, and $m_b/m_\tau$ resulting from mixing of states between this $X_{45}$ and the third generation matter ${\bf 16}$. 

The dominant sensitivity to $\alpha_s$ in this correlation arises from the determination of $v'$ from the running of the quartic, not from the determination of $v'$ from gauge coupling unification.  This implies that the sensitivity of the prediction for $\alpha_s$ to the grand unified thresholds, $\Delta_i$, as shown by the widths of the shadings in the left panel of Fig.~\ref{fig:3221_mta3}, is about an order of magnitude less in Higgs Parity unification than in conventional grand unification.

Taking the intermediate gauge symmetry to be 422 leads to a much larger value for $v'$ from gauge coupling unification: $v' \gsim 4 \times  10^{13}$ GeV, even allowing quite large unified threshold corrections, as shown in Fig.~\ref{fig:422_vp}.
To match the value of $v'$ from running of the SM quartic coupling then favors $x_Q$ values that successfully determine $m_b/m_\tau$, but only for large values of $\alpha_s$ and small values of $m_t$, as shown in Fig.~\ref{fig:422_mta3}.

In the 3221 theory with minimal content for the $SO(10)$ breaking Higgs, the unification scale is above $10^{16}$ GeV and the proton lifetime is predicted to be above the current constraint, as shown in the left panels of Fig.~\ref{fig:3221_vp}. An observation of proton decay at future experiments would require large threshold corrections at the unification scale, $\Delta \gsim 10$, and/or non-minimal $SO(10)$ breaking. In both cases, a larger $v'$ and hence a smaller top quark mass is favored, as illustrated in the right panel of Fig.~\ref{fig:3221_mta3}.

In the 422 theory, threshold corrections at the unification scale from $SO(10)$ breaking Higgses give $\Delta \sim O(1)$. As Fig.~\ref{fig:422_vp} shows, the theory predicts the unification scale around $10^{15}$ GeV and hence too short a proton lifetime. The unification scale can be raised to $10^{16}$~GeV by large threshold corrections, $\Delta>10$, which requires a rich structure around the unification scale such as $SO(10)$ symmetry breaking induced by supersymmetry breaking.

The observed flavor structure of the SM may arise from an $SO(10)$ unified theory, as suggested in Eq.~(\ref{eq:SO(10)yukawa}). Although we have not performed precise fits to the SM fermion masses, it would be interesting to do so and to investigate relations between the flavor observables. The model appears to predict a neutrino mass matrix proportional to the up quark mass matrix. However, this is avoided because of mixing between the third generation ${\bf 16}$ and ${\bf 45}/{\bf 54}$ fermions at the scale $v'$.  The theory of Eq.~(\ref{eq:SO(10)yukawa}) predicts $m_{\nu_{1.2}} \sim (v/v') m_{u,c}$, while the prediction is smaller by a factor of $(1/16 \pi^2)$ if $X_{45}$ is replaced by $X_{54}$; both cases give a normal neutrino mass hierarchy.  To obtain realistic neutrino masses requires $v' = (10^{10} \mathchar`- 10^{13} )$ GeV, which coincides with the scale required from gauge coupling unification and the vanishing SM quartic coupling. 
Because of the suppression, the yukawa coupling of the right-handed neutrinos responsible for the see-saw mechanism is larger than naively expected from the see-saw relation, increasing the efficiency of leptogenesis and allowing lower reheat temperatures than usual.

In conventional $SO(10)$ theories, the amount of fine tuning for symmetry breaking increases as the intermediate scale is reduced below the unification scale.  However, with Higgs Parity the amount of fine tuning is independent of the intermediate scale, and corresponds to the usual cost of keeping the weak scale below the cutoff.

\let\oldaddcontentsline\addcontentsline
\renewcommand{\addcontentsline}[3]{}

\section*{Acknowledgement}
This work was supported in part by the Director, Office of Science, Office of High Energy and Nuclear Physics, of the US Department of Energy under Contracts DE-AC02-05CH11231 (LH) and DE-SC0009988 (KH), as well as by the National Science Foundation under grants PHY-1316783 and PHY-1521446 (LH). 

\let\addcontentsline\oldaddcontentsline

\appendix

\section{Contributions of $X$ States to Beta Functions} 
\label{sec:Xcontrib}
In this Appendix we give the contributions of the $X$ states to the beta functions of the gauge couplings at two-loop level. We define the coefficient of the beta function by
\begin{align}
\frac{\rm d}{{\rm dln} \mu}
\begin{pmatrix}
\frac{2\pi}{\alpha_1} \\
\frac{2\pi}{\alpha_2} \\
\frac{2\pi}{\alpha_3}
\end{pmatrix} = 
\begin{pmatrix}
b_1 \\
b_2  \\
b_3
\end{pmatrix},~~
\frac{\rm d}{{\rm dln} \mu}
\begin{pmatrix}
\frac{2\pi}{\alpha_2} \\
\frac{2\pi}{\alpha_4}
\end{pmatrix} = 
\begin{pmatrix}
b_2  \\
b_4
\end{pmatrix}.
\end{align}
The contributions of each $X$ multiplet to the coefficients $b_i$
of the 3221 theory are
\begin{align}
X_{10} : \hspace{0.25in} &
\begin{pmatrix}
b_1 \\
b_2 \\
b_3
\end{pmatrix} = 
\begin{pmatrix}
- \frac{2}{3} \\
- \frac{2}{3}  \\
- \frac{2}{3}
\end{pmatrix} + 
\begin{pmatrix}
- \frac{1}{9} & 0  & - \frac{8}{9} \\
 0  & - \frac{29}{6} & 0 \\
-  \frac{1}{6} & 0 & - \frac{19}{3}
\end{pmatrix}
\begin{pmatrix}
\frac{\alpha_1}{2\pi} \\
\frac{\alpha_2}{2\pi} \\
\frac{\alpha_3}{2\pi}
\end{pmatrix}, \\
X_{45} : \hspace{0.25in} &
\begin{pmatrix}
b_1 \\
b_2 \\
b_3
\end{pmatrix} = 
\begin{pmatrix}
- \frac{16}{3} \\
-\frac{16}{3}  \\
- \frac{16}{3}
\end{pmatrix} + 
\begin{pmatrix}
- \frac{10}{3} & - 3  & - \frac{32}{3} \\
 - 1  & - \frac{119}{3} & -8 \\
- \frac{4}{3} & -6 & - \frac{167}{3}
\end{pmatrix}
\begin{pmatrix}
\frac{\alpha_1}{2\pi} \\
\frac{\alpha_2}{2\pi} \\
\frac{\alpha_3}{2\pi}
\end{pmatrix}, \\ 
X_{54} : \hspace{0.25in} &
\begin{pmatrix}
b_1 \\
b_2 \\
b_3
\end{pmatrix} = 
\begin{pmatrix}
- 8 \\
-8  \\
- 8
\end{pmatrix} + 
\begin{pmatrix}
- \frac{38}{9} & - 3  & - \frac{208}{9} \\
 - 1  & - \frac{131}{3} & -8 \\
- 4 & -6 & - 91
\end{pmatrix}
\begin{pmatrix}
\frac{\alpha_1}{2\pi} \\
\frac{\alpha_2}{2\pi} \\
\frac{\alpha_3}{2\pi}
\end{pmatrix},
\end{align}
and for the coefficients $b_{2,4}$ of the 422 theory are
\begin{align}
X_{10} : \hspace{0.25in} &
\begin{pmatrix}
b_2 \\
b_4
\end{pmatrix} = 
\begin{pmatrix}
- \frac{2}{3} \\
- \frac{2}{3}
\end{pmatrix} + 
\begin{pmatrix}
  - \frac{29}{6} &  0 \\
0 & - \frac{55}{6}
\end{pmatrix}
\begin{pmatrix}
\frac{\alpha_2}{2\pi} \\
\frac{\alpha_4}{2\pi}
\end{pmatrix}, \\
X_{45} : \hspace{0.25in} &
\begin{pmatrix}
b_2 \\
b_4
\end{pmatrix} = 
\begin{pmatrix}
- \frac{16}{3} \\
- \frac{16}{3}
\end{pmatrix} + 
\begin{pmatrix}
  - \frac{119}{3} &  - 15 \\
 -6 & - \frac{238}{3}
\end{pmatrix}
\begin{pmatrix}
\frac{\alpha_2}{2\pi} \\
\frac{\alpha_4}{2\pi}
\end{pmatrix}, \\
X_{54} : \hspace{0.25in} &
\begin{pmatrix}
b_2 \\
b_4
\end{pmatrix} = 
\begin{pmatrix}
- 8 \\
- 8
\end{pmatrix} + 
\begin{pmatrix}
  - 73 &  - 15 \\
 -6 & - \frac{221}{2}
\end{pmatrix}
\begin{pmatrix}
\frac{\alpha_2}{2\pi} \\
\frac{\alpha_4}{2\pi}
\end{pmatrix}.
\end{align}

\section{Threshold Corrections from $SO(10)$ Breaking Scalars}
\label{sec:SO10scalars}
In this Appendix we derive the threshold corrections to the gauge coupling unification from scalar multiplets that spontaneously break $SO(10)$.

\let\oldaddcontentsline\addcontentsline
\renewcommand{\addcontentsline}[3]{}

\subsection{$SU(3)\times SU(2)\times SU(2)\times U(1)$}
The smallest representation which can break $SO(10)$ down to $3221$ is ${\bf 45}$. This case is particularly interesting as the strong CP problem is solved by assigning an odd CP parity to ${\bf 45}$. The decomposition of ${\bf 45}$ into non-trivial $3221$ representations, and the contribution of each of these to the beta functions, is summarized in Table~\ref{tab:3221}.
The representations $(3,2,2,-1/3)$ and $(3,1,1,2/3)$ are would-be Nambu-Goldstone bosons. The threshold corrections to the gauge couplings are 
\begin{align}
\frac{2\pi}{\alpha_3(M_{XY})} = & \frac{2\pi}{\alpha_{10}(M_{XY})} - \frac{1}{2}{\rm ln} \frac{m_{(8,1,1)}}{M_{XY}} + \Delta_{3,G} \\
\frac{2\pi}{\alpha_2(M_{XY})} = & \frac{2\pi}{\alpha_{10}(M_{XY})} - \frac{1}{3} {\rm ln} \frac{m_{(1,3,1)}}{M_{XY}} + \Delta_{2,G} \\
\frac{2\pi}{\alpha_1(M_{XY})} = & \frac{2\pi}{\alpha_{10}(M_{XY})} + \Delta_{1,G}.
\end{align}
The contributions of ${\bf 45}$ to $\Delta_{ij} = \Delta_i - \Delta_j$ are 
\begin{align}
\Delta_{32}=  - \frac{1}{2}{\rm ln} \frac{m_{(8,1,1)}}{M_{XY}} +\frac{1}{3}{\rm ln} \frac{m_{(1,3,1)}}{M_{XY}}, ~~\Delta_{31} = - \frac{1}{2}{\rm ln} \frac{m_{(8,1,1)}}{M_{XY}},~\Delta_{21} =  -\frac{1}{3}{\rm ln} \frac{m_{(1,3,1)}}{M_{XY}}.
\end{align}
As shown in~\cite{Yasue:1980fy,Yasue:1980qj,Anastaze:1983zk,Bertolini:2009es}, after choosing the parameters of the potential to avoid tachyonic directions, $m_{8,1,1} = m_{1,3,1} = 0$. Their masses are given by quantum corrections, taking natural values of about $M_{XY}/10$. Even with this hierarchy, $\Delta_{ij}$ are only $ \approx 1$.

\begin{table}[htp]
\caption{Decomposition of ${\bf 45}$, ${\bf 54}$ and ${\bf 210}$ into representations of $3221$.}
\begin{center}
\begin{tabular}{|c|c|c|c|c|c|}
\hline
$SO(10)$ & \multicolumn{5}{c|}{{\bf 45}} \\ \hline
$SU(3)$ & 3 & 3 & 8  & 1 & 1 \\
$SU(2)$ & 2 & 1  & 1 & 3 & 1 \\
$SU(2)$ & 2 & 1 & 1 & 1 & 3 \\ 
$U(1)$ & $-1/3$ & 2/3 & 0 & 0 & 0 \\ \hline
$-b_3$  & 2/3 & 1/6 & 1/2 & \multicolumn{2}{c|}{0}  \\
$-b_2$ & 1 & 0 & 0 & \multicolumn{2}{c|}{1/3}  \\ 
$-b_{1}$ & 2/3 & 2/3 & 0 & \multicolumn{2}{c|}{0} \\ \hline
\end{tabular}
\begin{tabular}{|c|c|c|c|c|}
\hline
$SO(10)$ & \multicolumn{4}{c|}{{\bf 54}} \\ \hline
$SU(3)$ & 3 & 6 & 8  & 1 \\
$SU(2)$ & 2 & 1  & 1 & 3 \\
$SU(2)$ & 2 & 1 & 1 & 3  \\ 
$U(1)$ & $-1/3$ & $-2/3$ & 0 & 0  \\ \hline
$-b_3$  & 2/3 & 5/6 & 1/2 & 0  \\
$-b_2$ & 1 & 0 & 0 & 1  \\ 
$-b_{1}$ & 2/3 & 4/3 & 0 & 0  \\ \hline
\end{tabular}
\begin{tabular}{|c|c|c|c|c|c|c|c|c|c|c|c|c|c|}
\hline
$SO(10)$ & \multicolumn{12}{c|}{{\bf 210}} \\ \hline
$SU(3)$ & 3 & 3 & 8 & 1 &1 & 3 & 3 & 8 & 8 & 6 & 3 & 1  \\
$SU(2)$ & 2 & 1  & 1 & 3 & 1 & 3 & 1 & 3 & 1 & 2  & 2 & 2  \\
$SU(2)$ & 2 & 1  & 1  & 1 & 3  & 1 & 3 & 1 & 3 & 2 & 2 & 2 \\ 
$U(1)$ & $-1/3$ & 2/3 &0  & 0 & 0 & 2/3 & 2/3 & 0 & 0 & 1/3 & $-1/3$ & -1    \\ \hline
$-b_3$  & 2/3 & 1/6 & 1/2 & \multicolumn{2}{c|}{0}  &  \multicolumn{2}{c|}{1} &  \multicolumn{2}{c|}{3} & 10/3 & 2/3 & 0 \\
$-b_2$ & 1 & 0 & 0 & \multicolumn{2}{c|}{1/3}&  \multicolumn{2}{c|}{2} &  \multicolumn{2}{c|}{8/3} & 2 & 1 & 1/3 \\
$-b_1$ & 2/3 & 2/3 & 0 & \multicolumn{2}{c|}{0} &  \multicolumn{2}{c|}{4} &  \multicolumn{2}{c|}{0} & 4/3 & 2/3 & 2   \\ \hline
\end{tabular}
\end{center}
\label{tab:3221}
\end{table}%

We also consider ${\bf 54}$ whose decomposition is shown in Table~\ref{tab:3221}.  Although ${\bf 54}$ can break $SO(10)$ down only to $422$, its presence allows all components of ${\bf 45}$ to have positive mass squared at tree-level~\cite{Babu:1984mz}.  The threshold corrections from ${\bf 45}$ and ${\bf 54}$ are
\begin{align}
\begin{pmatrix}
\Delta_{32} \\
\Delta_{31} \\
\Delta_{21}
\end{pmatrix}
=
\begin{pmatrix}
-\frac{1}{2} & -\frac{1}{2} & \frac{1}{3} & \frac{1}{3} & - \frac{5}{6} & 1 \\
-\frac{1}{2} & -\frac{1}{2} & 0 & 0 & \frac{1}{2} & 0 \\
0 & 0 & - \frac{1}{3} & -\frac{1}{3} & \frac{4}{3} & -1
\end{pmatrix}
\begin{pmatrix}
{\rm ln} \frac{m_{(8,1,1)_1}}{M_{XY}} \\ 
{\rm ln} \frac{m_{(8,1,1)_2}}{M_{XY}} \\ 
{\rm ln} \frac{m_{(1,3,1)}}{M_{XY}} \\
{\rm ln} \frac{m_{(3,2,2)}}{M_{XY}} \\
{\rm ln} \frac{m_{(6,1,1)}}{M_{XY}} \\
{\rm ln} \frac{m_{(1,3,3)}}{M_{XY}} \\
\end{pmatrix}.
\end{align}
With $O(1)$ mass splittings, these threshold corrections can be $O(1)$.  With mass splittings of $O(10)$,  $\Delta$ can be $O(10)$; however such scalar mass hierarchies require fine-tuning of parameters.

We conclude that, in a theory with the strong CP problem solved by Higgs Parity, unified threshold corrections to gauge couplings are typically $O(1)$. However, threshold corrections can be large if the theory is non-minimal or the mass spectrum is fine-tuned, or if significant $SO(10)$ breaking feeds into the spectrum of $X$ states.

The next smallest representation is ${\bf 210}$, whose decomposition is shown in Table~\ref{tab:3221}. This representation is required if CP is not imposed on the theory. The contribution of ${\bf 210}$ to $\Delta_{ij}$ is
\begin{align}
\begin{pmatrix}
\Delta_{32} \\
\Delta_{31} \\
\Delta_{21}
\end{pmatrix}
=
\begin{pmatrix}
-\frac{1}{2} & \frac{1}{3} & 1 & -\frac{1}{3} & - \frac{4}{3} & \frac{1}{3} & \frac{1}{3} \\
-\frac{1}{2} & 0 & 3 & -3 & -2 & 0 & 2  \\
0 & -\frac{1}{3} & 2 & - \frac{8}{3} & - \frac{2}{3} & - \frac{1}{3} & \frac{5}{3}
\end{pmatrix}
\begin{pmatrix}
{\rm ln} \frac{m_{(8,1,1)}}{M_{XY}} \\ 
{\rm ln} \frac{m_{(1,3,1)}}{M_{XY}} \\
{\rm ln} \frac{m_{(3,3,1)}}{M_{XY}} \\
{\rm ln} \frac{m_{(8,3,1)}}{M_{XY}} \\
{\rm ln} \frac{m_{(6,2,2)}}{M_{XY}} \\
{\rm ln} \frac{m_{(3,2,2)}}{M_{XY}} \\
{\rm ln} \frac{m_{(1,2,2)}}{M_{XY}} 
\end{pmatrix}.
\end{align}
Depending on the mass spectrum, $\Delta$ may be as large as 10 even if the mass splittings are of $O(1)$.

\subsection{$SU(4)\times SU(2)\times SU(2)$}

\begin{table}[htp]
\caption{Decomposition of ${\bf 54}$ and ${\bf 210}$ into representations of $422$.}
\begin{center}
\begin{tabular}{|c|c|c|c|}
\hline
$SO(10)$ & \multicolumn{3}{c|}{{\bf 54}} \\ \hline
$SU(4)$ & 6 & $20'$ & 1  \\
$SU(2)$ & 2 & 1  & 3  \\
$SU(2)$ & 2 & 1 & 3  \\ \hline
$-b_4$  & 2/3 & 4/3 & 0  \\
$-b_2$ & 1 & 0 & 1  \\ \hline
\end{tabular}
\begin{tabular}{|c|c|c|c|c|c|c|}
\hline
$SO(10)$ & \multicolumn{5}{c|}{{\bf 210}} \\ \hline
$SU(4)$ & 6 & 15 & 15 & 15 & 10  \\
$SU(2)$ & 2 & 1   & 3  & 1 & 2  \\
$SU(2)$ & 2 & 1   & 1  & 3 & 2  \\ \hline
$-b_4$  & 2/3 & 2/3 & \multicolumn{2}{c|}{4} & 4  \\
$-b_2$ & 1 & 0 & \multicolumn{2}{c|}{5} & 10/3  \\ \hline
\end{tabular}
\end{center}
\label{tab:422}
\end{table}%

The smallest representation which can break $SO(10)$ down to $422$ is ${\bf 54}$. The decomposition of ${\bf 54}$ into the $422$ representations and the contribution of each to the beta functions are summarized in Table~\ref{tab:422}. The threshold corrections to the gauge couplings are 
\begin{align}
\frac{2\pi}{\alpha_2(M_{XY})} =& \frac{2\pi}{\alpha_{10}(M_{XY})} - {\rm ln} \frac{m_{(1,3,3)}}{M_{XY}} + \Delta_{2,G} \nonumber \\
\frac{2\pi}{\alpha_4(M_{XY})} = &\frac{2\pi}{\alpha_{10}(M_{XY})} - \frac{4}{3} {\rm ln} \frac{m_{(20',1,1)}}{M_{XY}} + \Delta_{4,G}.
\end{align}
Hence, the contribution of ${\bf 54}$ to $\Delta_{10}$ is 
\begin{align}
\Delta_{10,54} =   \frac{2}{3}{\rm ln} \frac{m_{(1,3,3)}}{M_{XY}} -  \frac{4}{3}{\rm ln} \frac{m_{(20',1,1)}}{M_{XY}},
\end{align}
which is a few at most, even if the mass splitting is $O(10)$.

The next smallest representation for breaking to 422 is ${\bf 210}$, whose decomposition is shown in Table~\ref{tab:422}. The strong CP problem is solved by assigning an odd CP parity to ${\bf 210}$.
The contribution of ${\bf 210}$ to $\Delta_{10}$ is 
\begin{align}
\Delta_{10,210} =  \frac{1}{3} {\rm ln} \frac{m_{(15,3,1)}^3 M_{XY}}{ m_{(15,1,1)}^2 m_{(10,2,2)}^2 } = {\rm ln} \frac{m_{(15,3,1)}}{ M_{XY}} - \frac{2}{3} \frac{m_{(15,1,1)}}{ M_{XY}} - \frac{2}{3} \frac{m_{(10,2,2)}}{ M_{XY}}
\end{align}
which is at most a few, even if the mass splitting is $O(10)$.

\let\addcontentsline\oldaddcontentsline

\bibliography{draft}

\end{document}